\documentclass[amsmath,amssymb,12pt,superscriptaddress,nofootinbib]{revtex4-1}

\usepackage{graphicx}
\usepackage{bm}
\usepackage{color}

\newcommand{\dd}{\mathrm{d}}

\newcommand{\del}{\partial}

\definecolor{DarkBlue}{rgb}{0,0,0.7}

\definecolor{DarkRed}{rgb}{0.65,0,0} 

\definecolor{DarkGreen}{rgb}{0,0.7,0} 
\newcommand{\dgreen}[1]{\textcolor{DarkGreen}{#1}}

\begin{document}
\baselineskip5.5mm
\thispagestyle{empty}


{\baselineskip0pt
\small
\leftline{\baselineskip16pt\sl\vbox to0pt{
               \hbox{\it Department of Physics, Rikkyo University} 
               \hbox{\it Department of Mathematics and Physics, Osaka City  University}
               \hbox{\it Division of Particle and Astrophysical Science, Nagoya University}
                             \vss}}
\rightline{\baselineskip16pt\rm\vbox to20pt{
            {
            \hbox{RUP-18-16}
            \hbox{OCU-PHYS-481}
            \hbox{AP-GR-146}
            }
\vss}}%
}

\author{Takahisa~Igata}\email{igata@rikkyo.ac.jp}
\affiliation{
Department of Physics, Rikkyo University, Toshima, Tokyo 171-8501, Japan
}

\author{Hideki~Ishihara}\email{ishihara@sci.osaka-cu.ac.jp}
\affiliation{ 
Department of Mathematics and Physics,Graduate School of Science, Osaka City University,
3-3-138 Sugimoto, Sumiyoshi, Osaka 558-8585, Japan
}

\author{Masataka~Tsuchiya}\email{tsuchiya.masataka@h.mbox.nagoya-u.ac.jp}
\affiliation{
Division of Particle and Astrophysical Science,
Graduate School of Science, Nagoya University, 
Nagoya 464-8602, Japan
}

\author{Chul-Moon~Yoo}\email{yoo@gravity.phys.nagoya-u.ac.jp}
\affiliation{
Division of Particle and Astrophysical Science,
Graduate School of Science, Nagoya University, 
Nagoya 464-8602, Japan
}

\vskip2cm

\title{
Rigidly Rotating 
String Sticking in a Kerr Black Hole
}

\begin{abstract}
\baselineskip5.5mm

\medskip
We analyze rigidly rotating Nambu--Goto strings in the Kerr spacetime, particularly
focusing on the strings sticking in the horizon.  From the
regularity on the horizon, we find the condition for sticking in the horizon, which is
consistent with the second law of the black hole thermodynamics.
Energy extraction through the sticking string from a Kerr black hole occurs.
We obtain the maximum value of the luminosity of the energy extraction. 
%
\end{abstract}

\maketitle
\pagebreak

%

\section{Introduction}

Motion of classical objects is one of the most fundamental 
blocks to construct a theory of physics. 
Needless to say, the motion of a test particle, 
which is described by a geodesic motion, plays an important role 
in the theoretical framework of general relativity. 
The geodesic motion can be also used to probe 
the geometric properties of a background curved spacetime. 
In this paper, we focus on another fundamental object, the Nambu--Goto string. 
This 
is the simplest $1+1$ object, whose
motion is governed by the action of the two-dimensional 
world sheet area. 
The classical motion of the string 
describes not only the motion of a macroscopic object like a cosmic string but also 
is used to analyze quantum aspects of a system through AdS/CFT correspondence
(see, e.g., Refs.~\cite{Lawrence:1993sg,Rey:1998ik,Maldacena:1998im}).

The motion of the Nambu--Goto string is generally described by 
a set of non-linear wave equations 
on its world sheet. 
Therefore, general string motion  
can hardly be analyzed without a highly sophisticated numerical procedure. 
Nevertheless, if the background spacetime admits a Killing vector field, 
and 
the string world sheet is foliated by the integral curves of the Killing vector field, 
we can reduce the equations of the string motion to 
a set of ordinary differential equations 
that describes a particle motion in  
the orbit space of the Killing vector field~\cite{Frolov:1988zn,Vilenkin:2000jqa}. 
We call these strings {\sl co-homogeneity-one strings}, c-1 strings for brevity.  
Classification and integrability of c-1 strings in maximally symmetric spacetimes were  
discussed in Refs.~\cite{Ishihara:2005nu,Koike:2008fs,Kozaki:2009jj,Morisawa:2017lpj}. 
As a class of c-1 strings, rigidly rotating string motions were studied 
in the Minkowski background~\cite{Frolov:1996xw,Ogawa:2008qn} and stationary black hole 
backgrounds~\cite{Frolov:1995vp,Frolov:1996xw,Kinoshita:2016lqd}. 
Gravitational perturbations sourced by a rigidly rotating string were analyzed 
in Ref.~\cite{Ogawa:2008yx}. 
A generalization of c-1 strings to c-1 membranes was 
given in Ref.~\cite{Kozaki:2014aaa}.

In particular, the system of 
a rigidly rotating string in a Kerr black hole is 
worthy of attention. 
As is reported in Ref.~\cite{Kinoshita:2016lqd}(see also Ref.~\cite{Kinoshita:2017mio}), 
energy extraction from a Kerr black hole is possible 
through a stationary rotating string  
analogously to the widely known extraction processes: super-radiance~\cite{1971JETPL..14..180Z,1972JETP...35.1085Z,Starobinsky:1973aij,1974JETP...38....1S,Teukolsky:1974yv,1972ApJ...178..347B}, 
the Penrose process~\cite{Penrose:1969pc,Penrose:1971uk}, 
and the Blandford--Znajek process~\cite{Blandford:1977ds}. 
Furthermore, 
the system is interesting if the classical string motion can be 
related with some quantum aspects of field theories 
in the context of the Kerr/CFT correspondence~\cite{Guica:2008mu}.

In this paper, we clarify the natural question: whether a rigidly rotating string 
can stick into the black hole horizon or not. 
In the paper by Frolov et al.~\cite{Frolov:1996xw}, a stationary string winds infinite times around 
a black hole just outside its horizon. We obtain the condition for a string to stick into 
the horizon by analyzing the equations of motion in the Kerr coordinates, which regularly cover the horizon.

This paper is organized as follows. 
In Sec.~\ref{sec:eom}, we derive the reduced equations of motion for a rigidly rotating Nambu--Goto string  
in the Kerr spacetime. 
We show that the allowed region for the motion of the string is bounded by the light surface and 
the centrifugal barrier surface. 
The configuration of the allowed region is classified in Sec.~\ref{sec:class} 
for every parameter set, and we discuss the regularity on the light surface, 
where the norm of the associated Killing vector becomes null.  
In Sec.~\ref{sec:reghori}, 
we show that 
the string can regularly penetrate through the horizon, 
depending on the direction of its physical energy flux, 
and find it consistent with the area law of the black hole thermodynamics. 
The maximum value of the luminosity of the energy extraction 
through a rigidly rotating string from a Kerr black hole 
is given in Sec.~\ref{sec:maxlumi}. 
Section~\ref{sec:sumdis} is devoted to a summary and discussion.

In this paper, we use geometrized units in which both
the speed of light $c$ and Newton's gravitational constant $G$ are 
one.

\section{Equations of motion 
}
\label{sec:eom}

The line element of the Kerr spacetime in the Boyer--Lindquist coordinates is given by
\begin{equation}
	\dd s^2=g_{\mu\nu}\dd x^\mu \dd x^\nu=-\left(1-\frac{2Mr}{\Sigma}\right)\dd t^2
		-\frac{4Mar}{\Sigma}\sin^2\theta \dd t\dd \bar \phi 
		+\frac{A}{\Sigma}\sin^2\theta\dd \bar \phi^2
		+\frac{\Sigma}{\Delta}\dd r^2+\Sigma \dd \theta^2 
\end{equation}
with
\begin{align}
	\Delta(r)&:=r^2-2Mr+a^2, \\
	\Sigma(r,\theta)&:=r^2+a^2\cos^2\theta,\\
	A(r,\theta)&:=\left(r^2+a^2\right)^2-\Delta a^2\sin^2\theta,  
\end{align}
where $M$ and $a$ are the mass of the black hole and the spin parameter 
satisfying $0\leq|a|\leq M$. 
The radius $r_+$  of the outer horizon is given by
\begin{equation}
	r_+=M+\sqrt{M^2-a^2}. 
\end{equation}

We consider rigidly rotating Nambu--Goto strings, that is, 
a class of c-1 strings associated with the Killing vector $\xi$ given by
\begin{equation}
	\xi=\del_t+\Omega \del_{\bar\phi},  
\label{Killing_vec}
\end{equation}
where the constant $\Omega$ is the angular velocity. 
Such a kind of Nambu--Goto string configuration is obtained by solving 
geodesic equations on the reduced space 
with the metric \cite{Frolov:1988zn,Vilenkin:2000jqa} 
\begin{equation}
	\tilde h_{ij}\dd x^i\dd x^j
		:=-\left(fg_{\mu\nu}-\xi_\mu\xi_\nu\right)\dd x^\mu \dd x^\nu
		=-f\left(\frac{\Sigma}{\Delta}\dd r^2+\Sigma\dd\theta^2\right)
			+\Delta \sin^2\theta \dd \phi^2, 
\end{equation}
where $\phi=\bar \phi-\Omega t$. 
The function $f(r,\theta)$ is the norm of $\xi$: 
\begin{equation}
	f(r,\theta):=g_{\mu\nu}\xi^\mu\xi^\nu=-1+\frac{2Mr}{\Sigma}
		-\frac{4Mra\sin^2\theta}{\Sigma}\Omega+\frac{A\sin^2\theta}{\Sigma}\Omega^2.
\label{f}
\end{equation}
The world sheet of the c-1 string is obtained by the foliation of integral curves 
of the Killing vector field $\xi$ along the geodesic curve.
The spacetime outside the outer horizon is divided by the surface $f=0$, 
we call it {\sl\lq light surface\rq }.  
The rigidly rotating strings in the region where $\xi$ is timelike, $f<0$, 
are stationary rotating strings, 
while the strings in the region where $\xi$ is spacelike, $f>0$, 
are non-stationary rotating strings. 
Since the relative sign of $\Omega$ to $a$ is relevant in 
the following analyses,  
we assume $a\geq 0$ in this paper. 

The action for a geodesic particle in the reduced space  
can be written in the form
\begin{equation}
	\mathcal S=\frac{1}{2}\int\left(\frac{1}{N}\tilde h_{ij}\dot x^i\dot x^j+N\right)
	\dd \sigma, 
\label{geodesic_action}
\end{equation}
where the dot \lq$\dot ~$\rq\ denotes the derivative with respect to 
a parameter $\sigma$ on the world line, 
and $N$ is the Lagrange multiplier.  
From the action \eqref{geodesic_action}, we obtain the Hamiltonian 
in the form 
\begin{equation}
	\mathcal H
		=\frac{N}{2}\left(\tilde h^{ij}p_ip_j-1\right)
		=\frac{N}{2}\left(-\frac{\Delta}{f\Sigma}p_r^2-\frac{1}{f\Sigma}p_\theta^2
			+ \frac{p_\phi^2}{\Delta\sin^2\theta}-1\right),
\end{equation}
where $p_i:=N^{-1} \tilde h_{ij} \dot x^j$ is the conjugate momentum to $x^i$.

Setting the Lagrange multiplier $N$ as 
\begin{equation}
	N=-f\Sigma ,  
\label{laps}
\end{equation}
we obtain the equations of motion as 
\begin{align}
\dot r&=\Delta p_r, 
\label{eom_r}\\
	\dot p_r&=
		-\frac{1}{2}\left(\del_r \Delta p_r^2+\del_r V\right), 
\label{eom_pr}\\
	\dot \theta&=p_\theta, 
\label{eom_th}\\
	\dot p_\theta&=-\frac{1}{2}\del_\theta V, 
\label{eom_pth}\\
	\dot \phi&=\frac{qf\Sigma}{\Delta\sin^2\theta}. 
\label{eom_ph}
\end{align}
Since $\phi$ is a cyclic coordinate,
we can set 
\begin{align}
	p_\phi=-q={\rm const}. 
\label{q}
\end{align}

Further, the Hamiltonian constraint, 
obtained by the variation of $N$, is written as 
\begin{equation}
	\mathcal H=\Delta p_r^2+p_\theta^2+V \approx 0 ,
\label{constr}
\end{equation}
where 
\begin{align}
	&V := f	k\Sigma, \\ 
	&k(r, \theta) :=1-\frac{q^2}{\Delta\sin^2\theta}. 
\label{k}
\end{align}
We call the surface $k(r, \theta)=0$ the centrifugal barrier because $k$
is characterized by $q=-p_\phi$.
Since $\Delta$ is non-negative 
outside the horizon, then $V$ in Eq.~\eqref{constr} 
must be non-positive. Since  
$\Sigma$ is also positive, 
the allowed region of the string motion is given by 
\begin{equation}
	f(r,\theta)k(r,\theta)\leq 0.  
\label{allowed}
\end{equation}

\section{Classification 
of the allowed region} 
\label{sec:class}

We consider the spatial configuration of the light surface $f=0$. 
It is obvious that the light surface has axial symmetry and reflection symmetry 
to the equatorial plane. 
Therefore, we pay attention to the curve $f(r,\theta)=0$ on the $r$-$\theta$ plane. 
The topology of the light surface is classified by the number of intersection of 
$f=0$ and $\theta=\pi/2$, the  equatorial plane. Then, we consider the equation 
\begin{equation}
	f\left(r,\frac{\pi}{2}\right)=\frac{1}{r}\left[\Omega^2r^3
		-\left(1-a^2\Omega^2\right)r+2M\left(1-a\Omega\right)^2\right]=0 . 
\label{eq:X}
\end{equation}
As is shown in Appendix \ref{sec:Xr0}, 
the number of roots of Eq.~\eqref{eq:X} 
in the region $r> r_+$ for $0\leq a\leq M$ is summarized as follows: 
\begin{align}
	2~{\rm roots} \quad {\rm for}&\quad  
	\Omega^-_{\rm cr}<\Omega  < \Omega^+_{\rm cr},
\label{eq:tworoot}\\
	1~{\rm roots} \quad {\rm for}&\quad \Omega = \Omega^\pm_{\rm cr},
\label{eq:oneroot}\\
	0~{\rm roots} \quad {\rm for}&\quad 
		\Omega<\Omega^-_{\rm cr}~~{\rm or}~~ \Omega^+_{\rm cr}<\Omega , 
\end{align}
where $\Omega^\pm_{\rm cr}$ are positive and negative roots of the equation 
\begin{equation}
	27(M\Omega_{\rm cr})^2 \left(1-a\Omega_{\rm cr}\right)
		=\left(1+a\Omega_{\rm cr}\right)^3 
\label{eq:forcr}
\end{equation}
in the range of $|a\Omega^\pm_{\rm cr}|<1/2$. The explicit forms of $\Omega^\pm_{\rm cr}$ are 
given in Appendix~\ref{sec:Xr0}. 

\begin{figure}[!htbp]
\begin{center}
\includegraphics[scale=0.7]{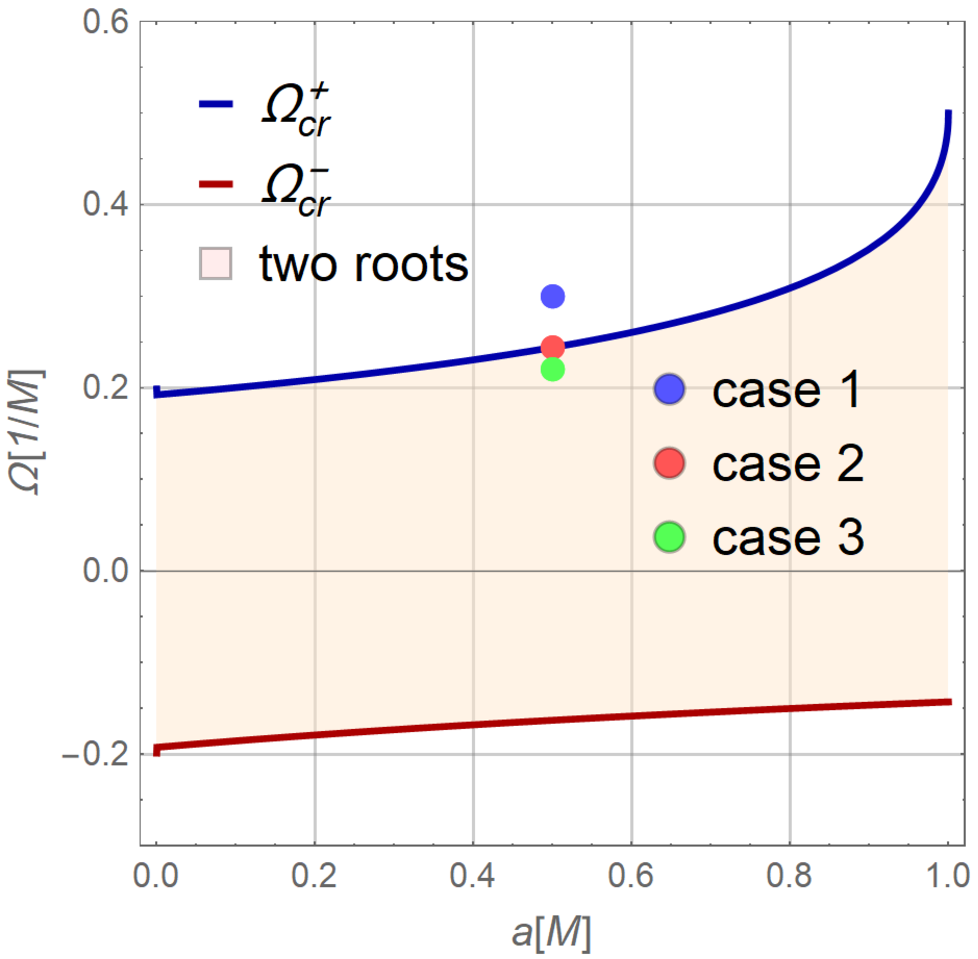}\qquad
\includegraphics[scale=0.7]{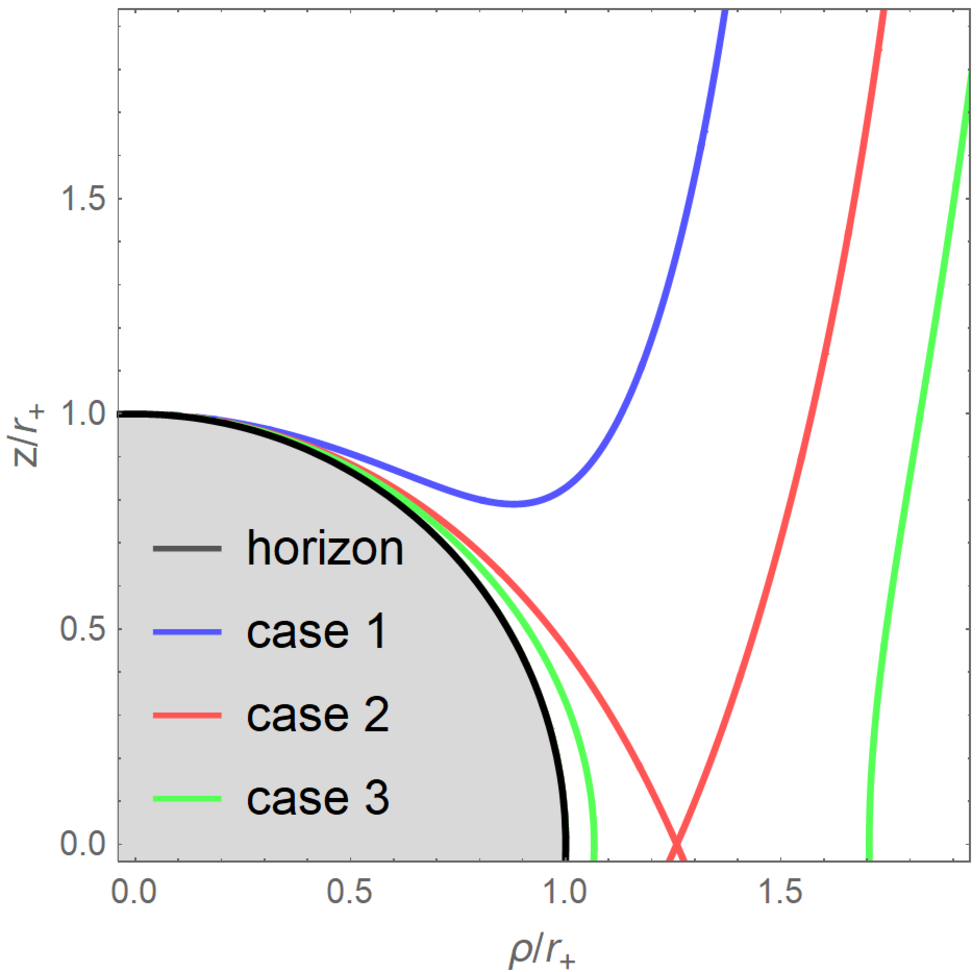}
\caption{\baselineskip5.5mm
Parameter regions and the number of roots for Eq.~\eqref{eq:X}~(left).  
Some examples of the light surfaces on a $t={\rm const.} ~\phi={\rm const.}$ plane~(right), 
where $z:=r\cos\theta$ and $\rho:=r\sin\theta$. 
}
\label{fig:solnum}
\end{center}
\end{figure}

The parameter $\Omega$ characterizes the shape of the light surface as shown 
in Fig.~\ref{fig:solnum}. 
For a fixed $a$ in the range $|a|<M$, 
if $\Omega^-_{\rm cr}<\Omega<\Omega^+_{\rm cr}$, a cylindrical light surface 
and a spherical light surface appear, 
and two surfaces move closer to each other as $\Omega$ approaches to the critical 
values $\Omega^{\pm}_{\rm cr}$. 
If $\Omega=\Omega^\pm_{\rm cr}$, these surfaces touch each other at a point, say X, 
on the equatorial plane, 
and if $\Omega<\Omega^-_{\rm cr}$ or $\Omega^+_{\rm cr}<\Omega$,  
the two light surfaces merge and change to a bottle-like surface 
on the 
pole of the horizon, 
respectively (see Fig.~\ref{fig:solnum}).

We evaluate $f(r,\theta)$ on the horizon as
\begin{equation}
	f(r_+,\theta)
	=\frac{(r_+^2+a^2)^2\sin^2\theta}{\Sigma(r_+,\theta)}
		\left(\Omega-\Omega_{\rm H}\right)^2 \geq 0, 
\end{equation}
where 
\begin{align}
	\Omega_{\rm H}:=\frac{a}{r_+^2+a^2}
\end{align} 
is the so-called horizon angular velocity. 
We see that the horizon is surrounded by the region $f>0$ for $\Omega\neq \Omega_{\rm H}$, and the light surface touches the horizon at the north and south poles. 
If $\Omega=\Omega_{\rm H}$, the spherical light surface 
coincides with the horizon as the exceptional case.

\subsection{Intersection between the light surface and the centrifugal barrier}
The centrifugal barrier $k=0$ in the Kerr geometry is a cylindrical surface. 
The configuration of the allowed region $f k \leq 0$ of the rigidly rotating strings 
is classified 
by inspecting the intersection of the light surface and the centrifugal barrier, or 
equivalently, the intersecting point S of curves 
$f(r,\theta)=0$ and $k(r,\theta)=0$ on the $r$-$\theta$ plane. 
Because of the reflection symmetry with respect to the equatorial plane, 
we consider the position of the point S, $(r_{\rm S}, \theta_{\rm S})$, in the range 
$0 \leq \theta_{\rm S} \leq \pi/2$.

We depict the parameter regions of $q$ and $\Omega$ which allow the intersection between 
the centrifugal barrier and the light surface in Fig.~\ref{fig:pararegion}. 
A simple derivation of the boundary curves of the allowed region is given in Appendix~\ref{sec:irrex} and 
detailed analyses on the shape of the parameter region are given in Appendix~\ref{sec:irrex_det}. 
For some parameter sets of $q$ and $\Omega$ marked 
in the right panel in Fig.~\ref{fig:pararegion}, the corresponding allowed regions are shown 
in Fig.~\ref{fig:cases}.

For a fixed $\Omega$ in the range $\Omega^-_{\rm cr}\leq \Omega \leq \Omega^+_{\rm cr}$, 
where 
light surface consists of a spherical surface and a cylindrical surface, 
if $|q|=0$, the curve $k=0$ intersects with the spherical light surface at the north 
pole. 
As $|q|$ increases, $r_{\rm S}$ and $\theta_{\rm S}$ increase along the spherical light surface, 
and the point S reaches the equatorial plane, $\theta_{\rm S}=\pi/2$, 
and then the intersection disappears. 
Increasing $|q|$ further, we see that $k=0$ 
touches the cylindrical light surface  
on the equatorial plane, $\theta_{\rm S}=\pi/2$, and $r_{\rm S}$ increases and 
$\theta_{\rm S}$ decreases along the cylindrical light surface. 
Finally, the point S goes as $r_{\rm S} \to \infty$ and $\theta_{\rm S}\to 0$ 
at a critical value of $|q|$.

For a fixed $\Omega$ in the range $\Omega<\Omega^-_{\rm cr}$ or $\Omega^+_{\rm cr}<\Omega $, 
where the light surface has a bottle-like shape, 
as $|q|$ increases from $0$, the point S moves from the north pole to infinity, 
i.e., $(r_{\rm S}= r_+, \theta_{\rm S}=0) \to (r_{\rm S}= \infty, \theta_{\rm S}=0)$.  
In the special case $\Omega=\Omega^\pm_{\rm cr}$, 
for 
special values of $q_\pm$ in Fig.~\ref{fig:pararegion}, 
the point S coincides with the point X on the equatorial plane. 
The intersecting point S appears for limited regions 
on the parameter space $(\Omega, q)$. 
As is shown in Appendix \ref{sec:irrex}, the regions are bounded by the curves:
\begin{align}
	\Omega^2 q^2=1, 
\end{align}
and
\begin{align}
	&\Omega^4 q^6 
		-\Omega^2 \left(a^2 \omega^2+2\right)q^4 
\cr
	&-\left(a^2 \left(8 M^2 \Omega^4-2 \Omega^2\right)-24 a M^2 \Omega^3
		+16  M^2 \Omega^2-1\right)q^2 
\cr
	&-\left(4 a M^2 \Omega^2+a-4 M^2 \Omega \right)^2 =0.
\label{boundary_curve_2}
\end{align}

The region for the existence of S is divided by the point 
$(\Omega, q)=(\Omega_H, 0)$ in two parts, 
and the radius of S in the 
right region is given by 
\begin{align}
	r_{\rm S}&=\frac{M}{1- |q|\Omega}
		+\sqrt{\left(\frac{M}{1- |q|\Omega}\right)^2-a^2- |q| a}, 
\end{align}
and the same in the 
left region is given by
\begin{align}
	r_{\rm S}=\frac{M}{1 + |q|\Omega}
		+\sqrt{\left(\frac{M}{1 + |q|\Omega}\right)^2-a^2 +|q| a} ~. 
\end{align}
The value $\theta_{\rm S}$ is obtained by $k(r_{\rm S}, \theta_{\rm S})=0$ as 
\begin{align}
	\sin\theta_{\rm S}=\frac{q}{\sqrt{\Delta(r_{\rm S})}}.
\end{align} 
Typical configurations of the centrifugal barrier and the light surfaces are 
shown in Fig.~\ref{fig:cases}. 

\begin{figure}[!htbp]
\begin{center}
\includegraphics[scale=0.85]{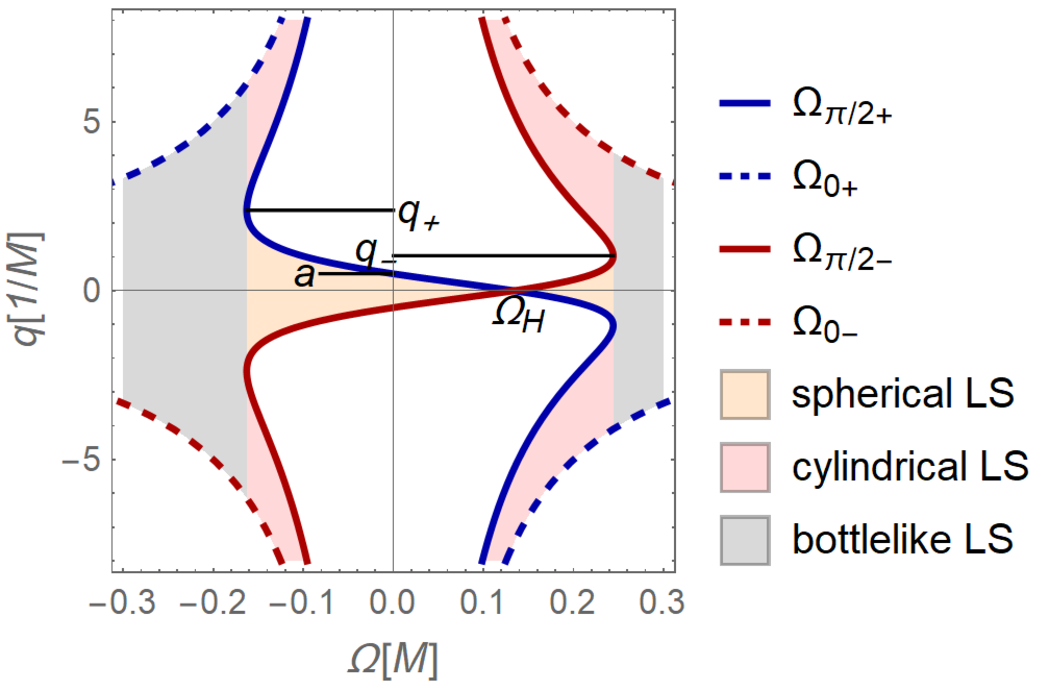}
\includegraphics[scale=0.65]{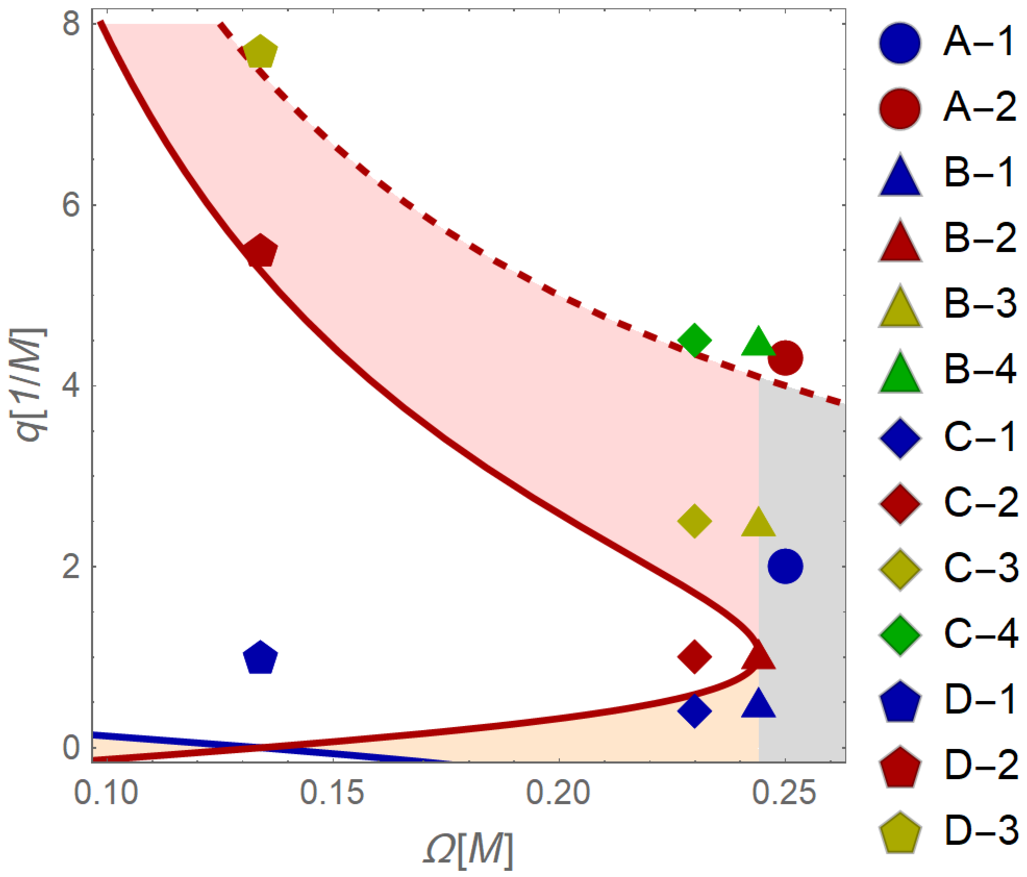}
\caption{\baselineskip5.5mm
The parameter regions which allow the intersection between the 
centrifugal barrier and the light surface are depicted for $a=M/2$(left panel),
where ``LS" is the abbreviation of light surface.  
The parameter region can be classified into three regions 
by the configuration of the light surface which intersects with the 
centrifugal barrier. 
The right panel is the scale-up figure of the upper right part of the left figure 
with the corresponding points to the configurations depicted in Fig.~\ref{fig:cases}. 
}
\label{fig:pararegion}
\end{center}
\end{figure}

%
\begin{figure}[!ht]
\begin{center}
\includegraphics[scale=0.3]{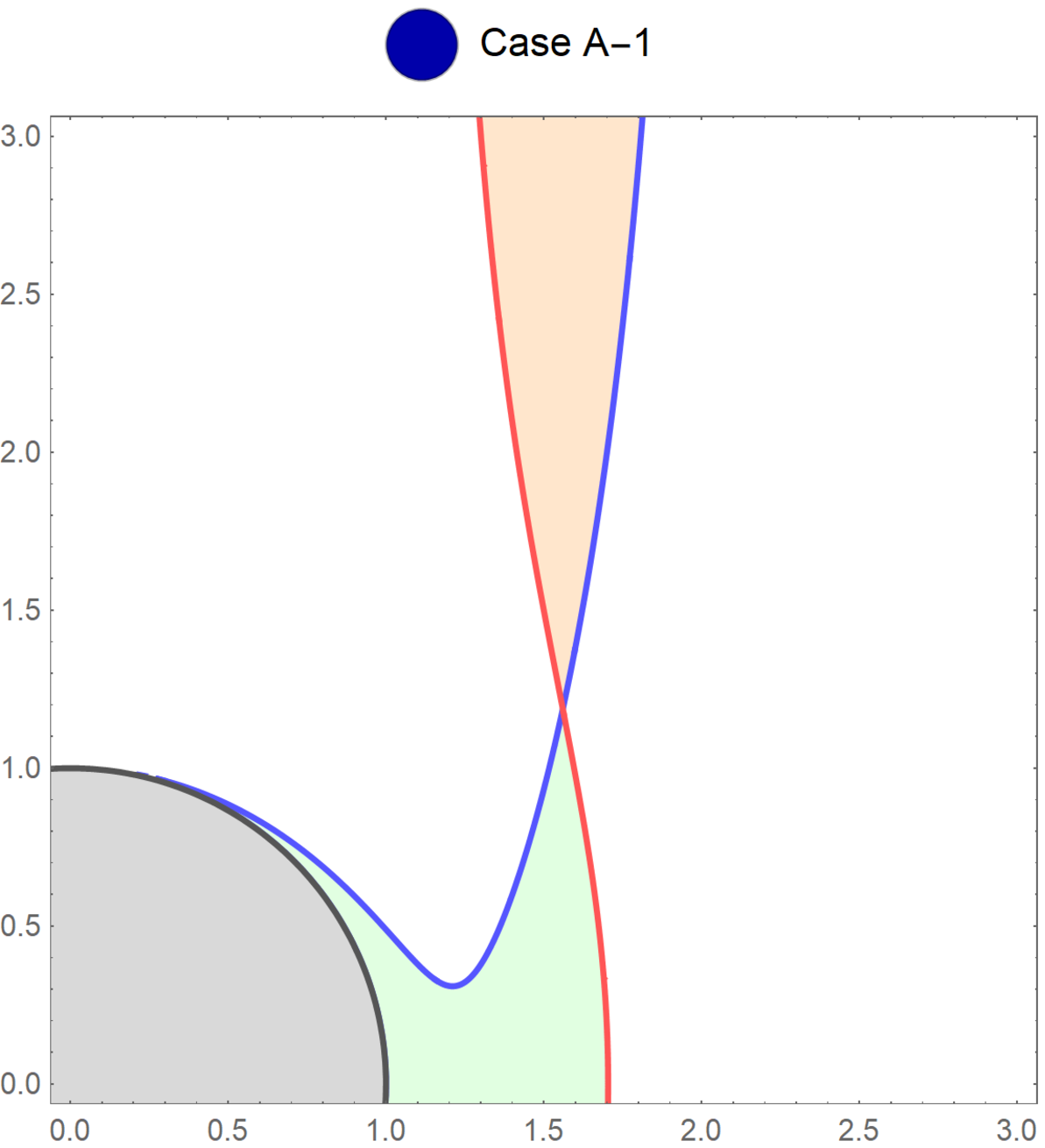}
\includegraphics[scale=0.3]{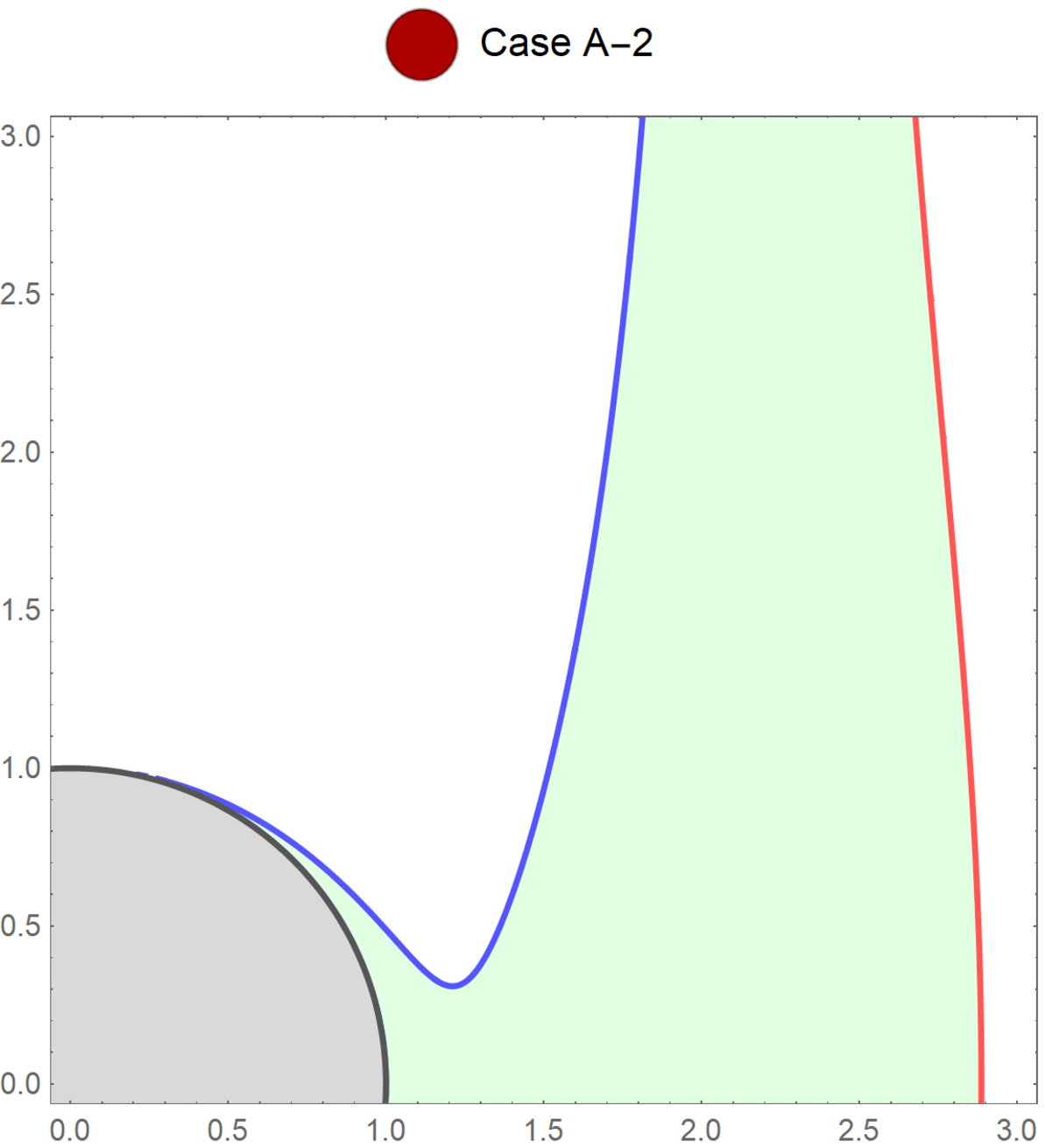}
\includegraphics[scale=1.]{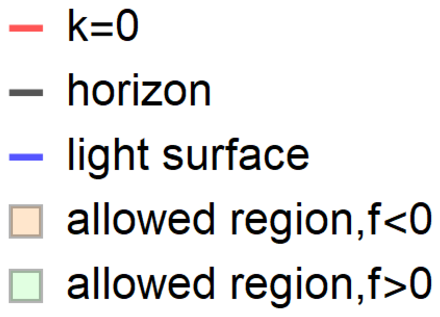}\\
\vspace{7mm}
\includegraphics[scale=0.3]{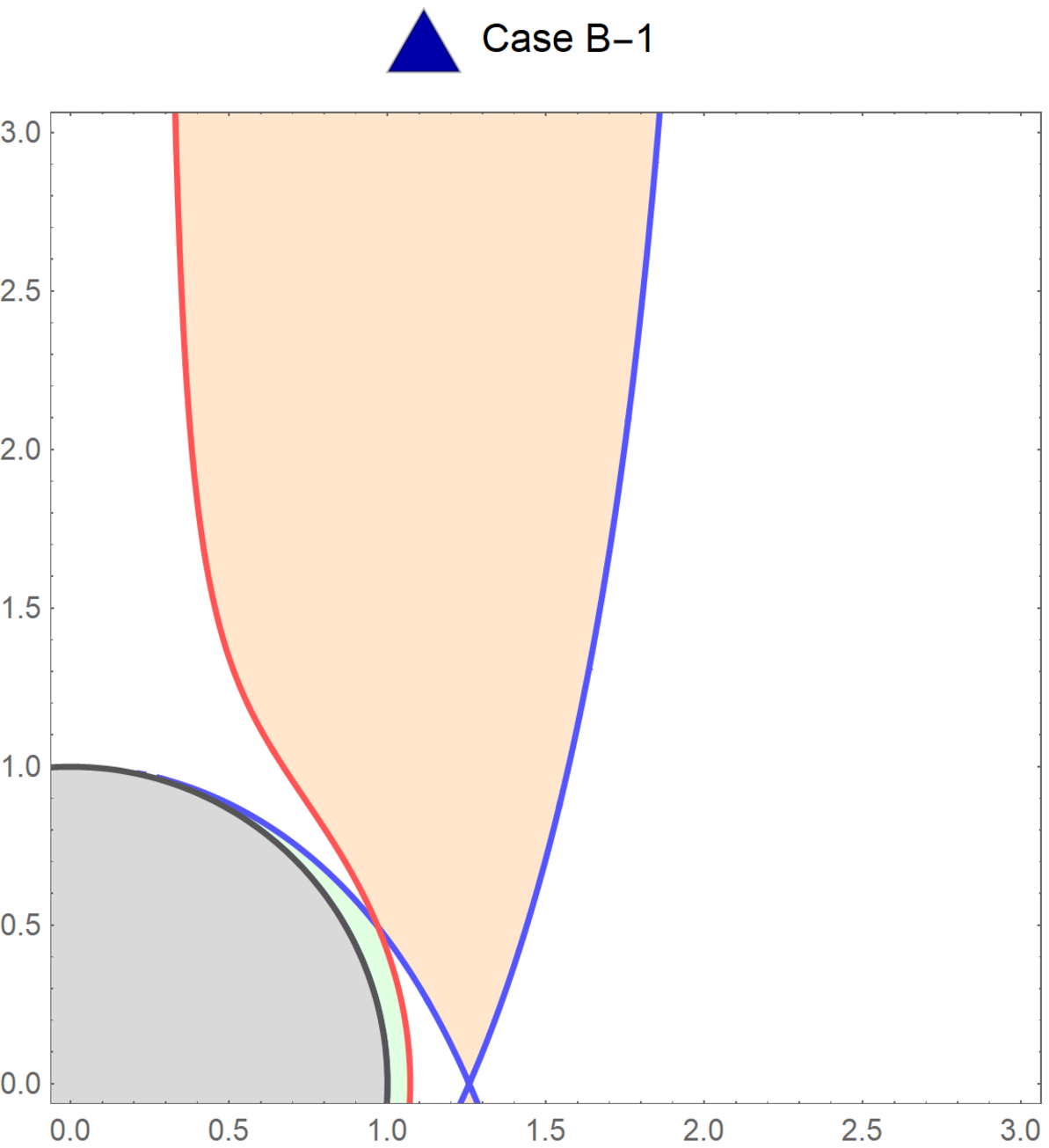}
\includegraphics[scale=0.3]{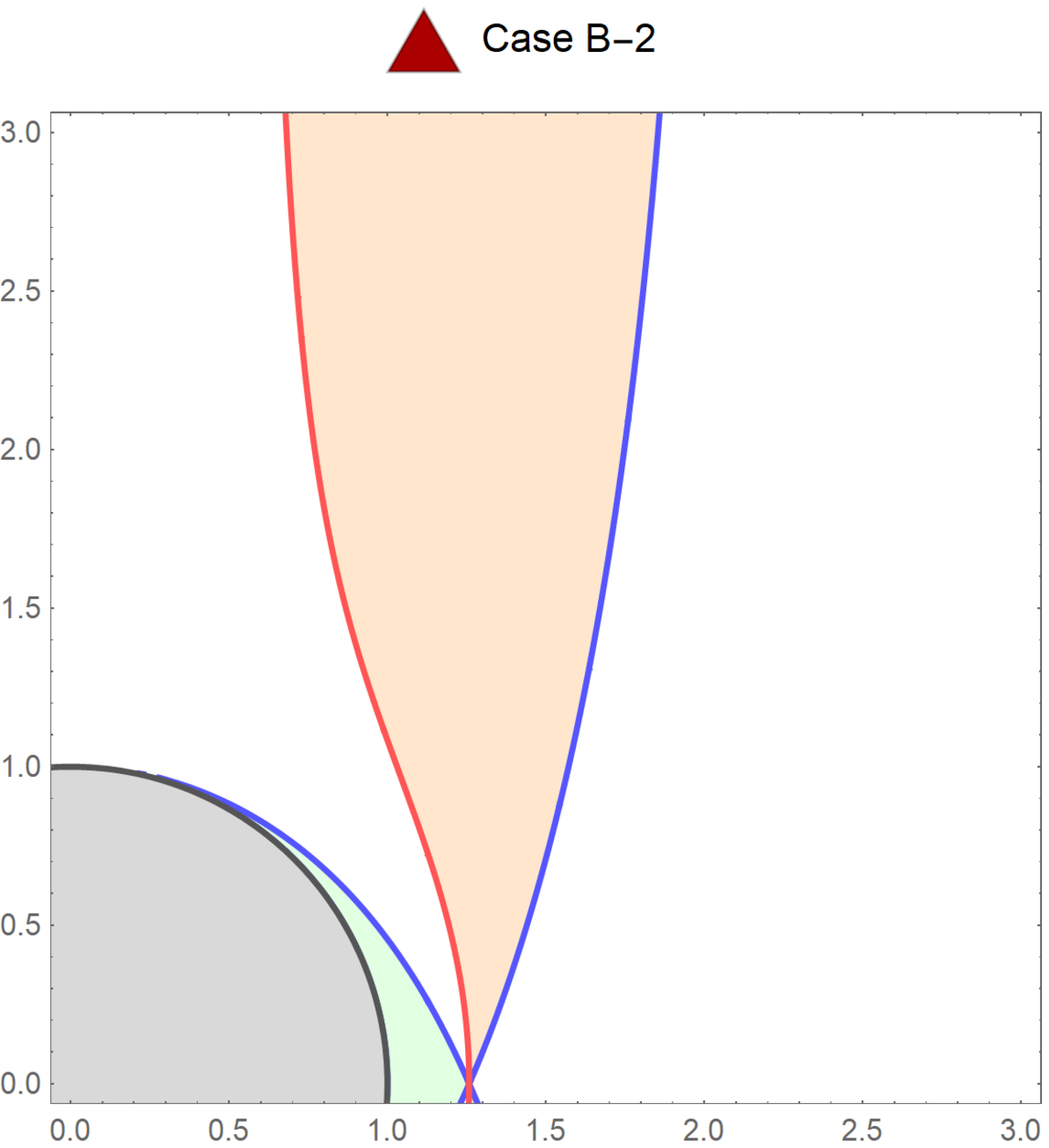}
\includegraphics[scale=0.3]{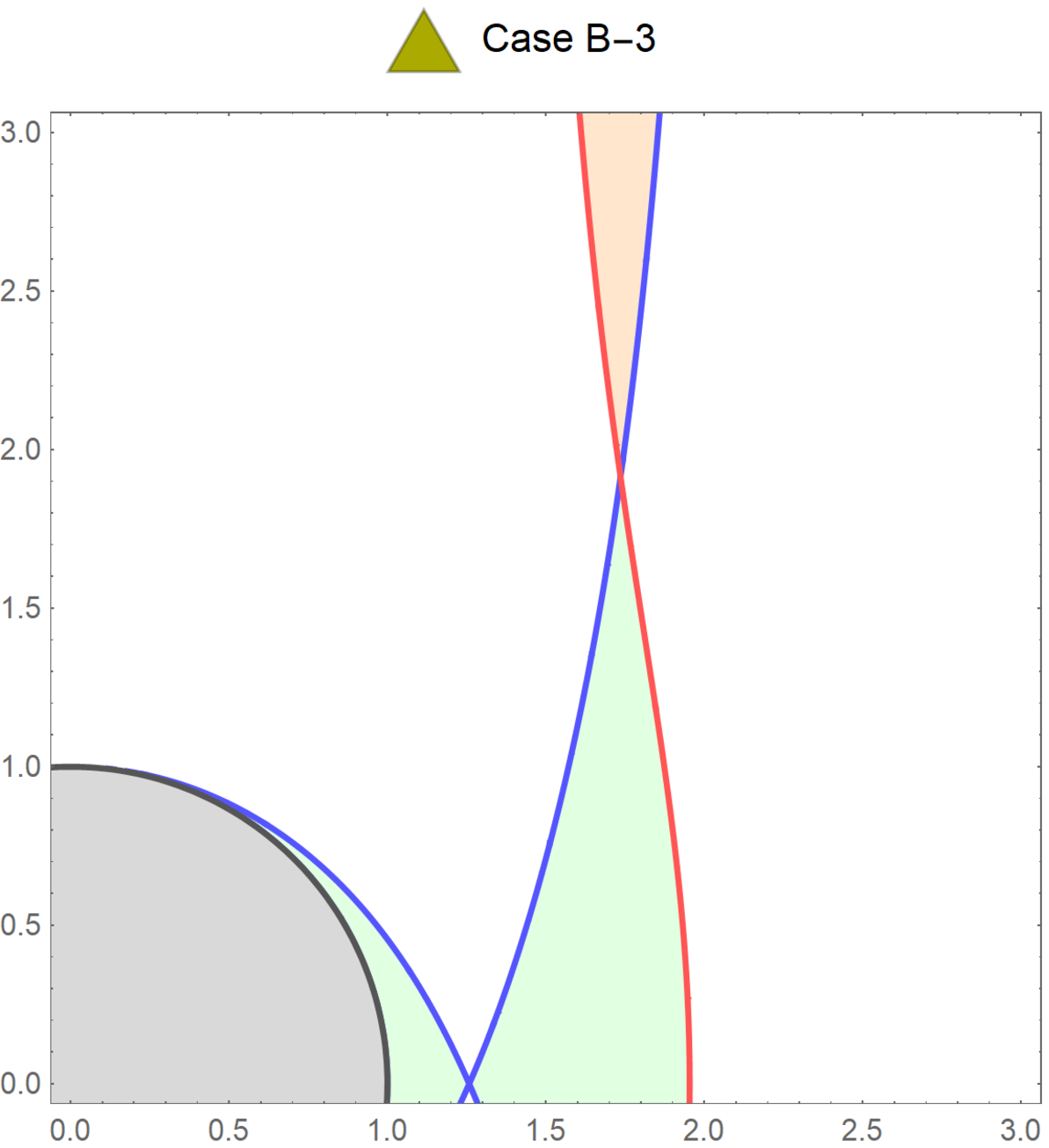}
\includegraphics[scale=0.3]{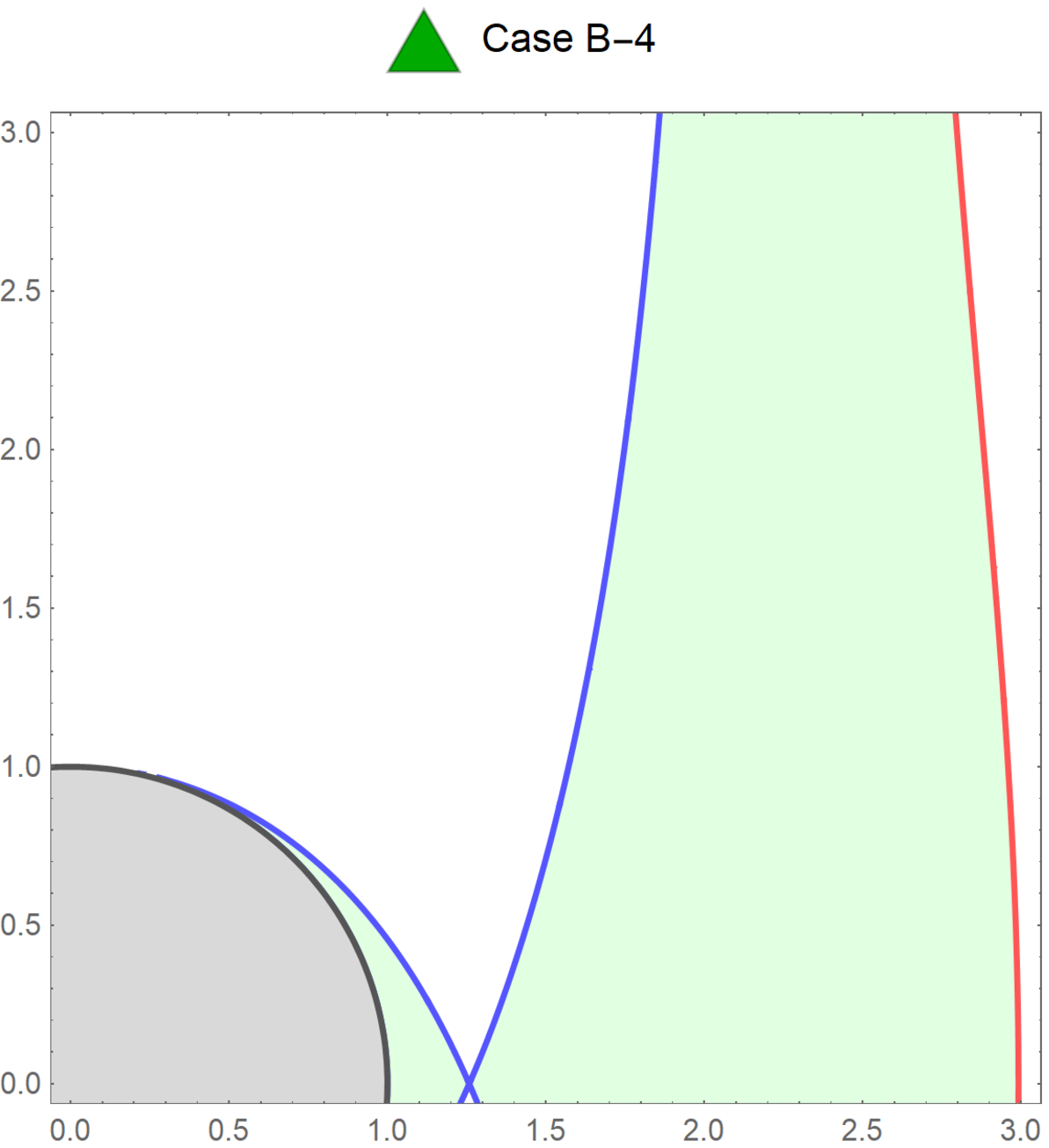}\\
\vspace{7mm}
\includegraphics[scale=0.3]{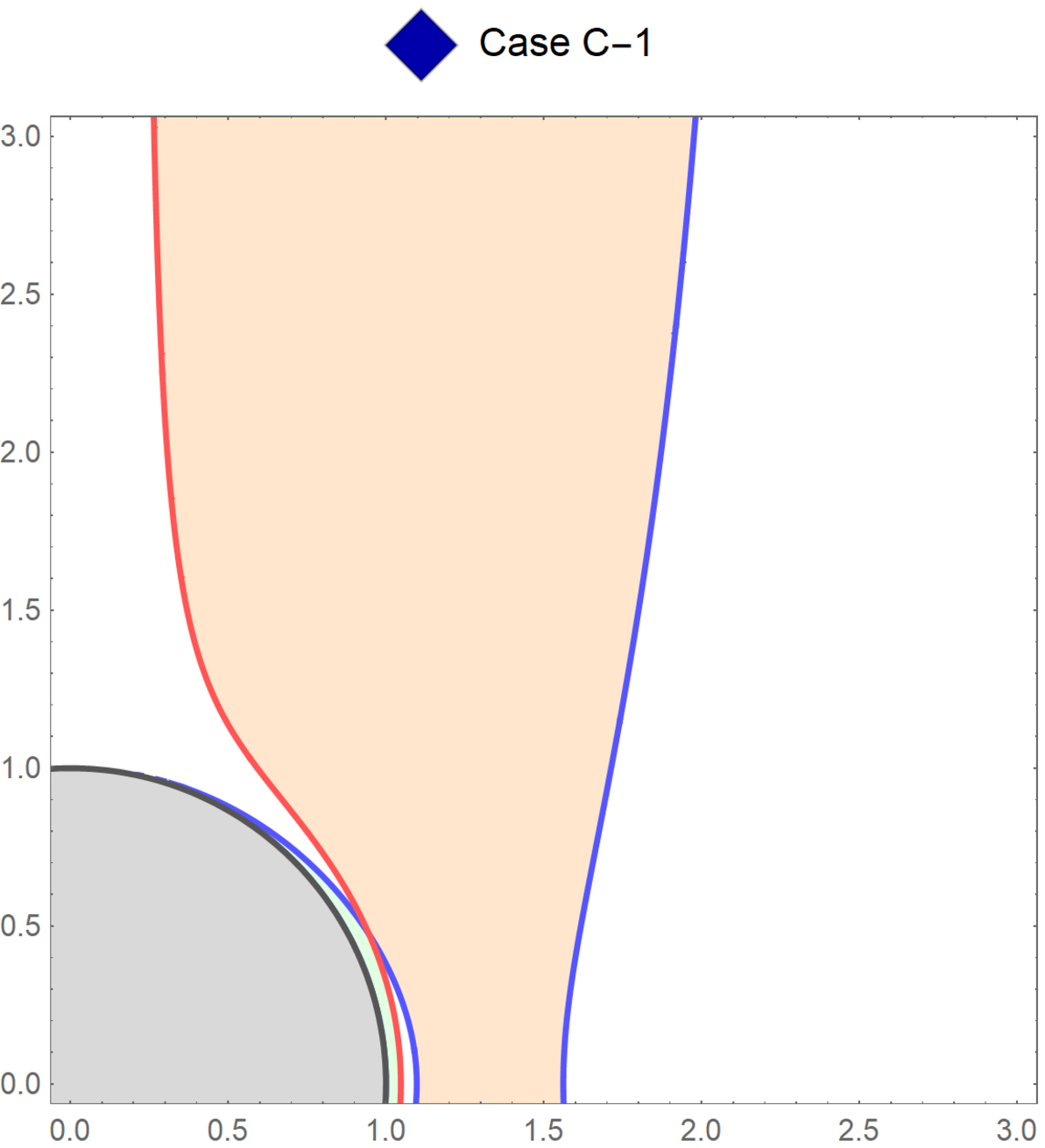}
\includegraphics[scale=0.3]{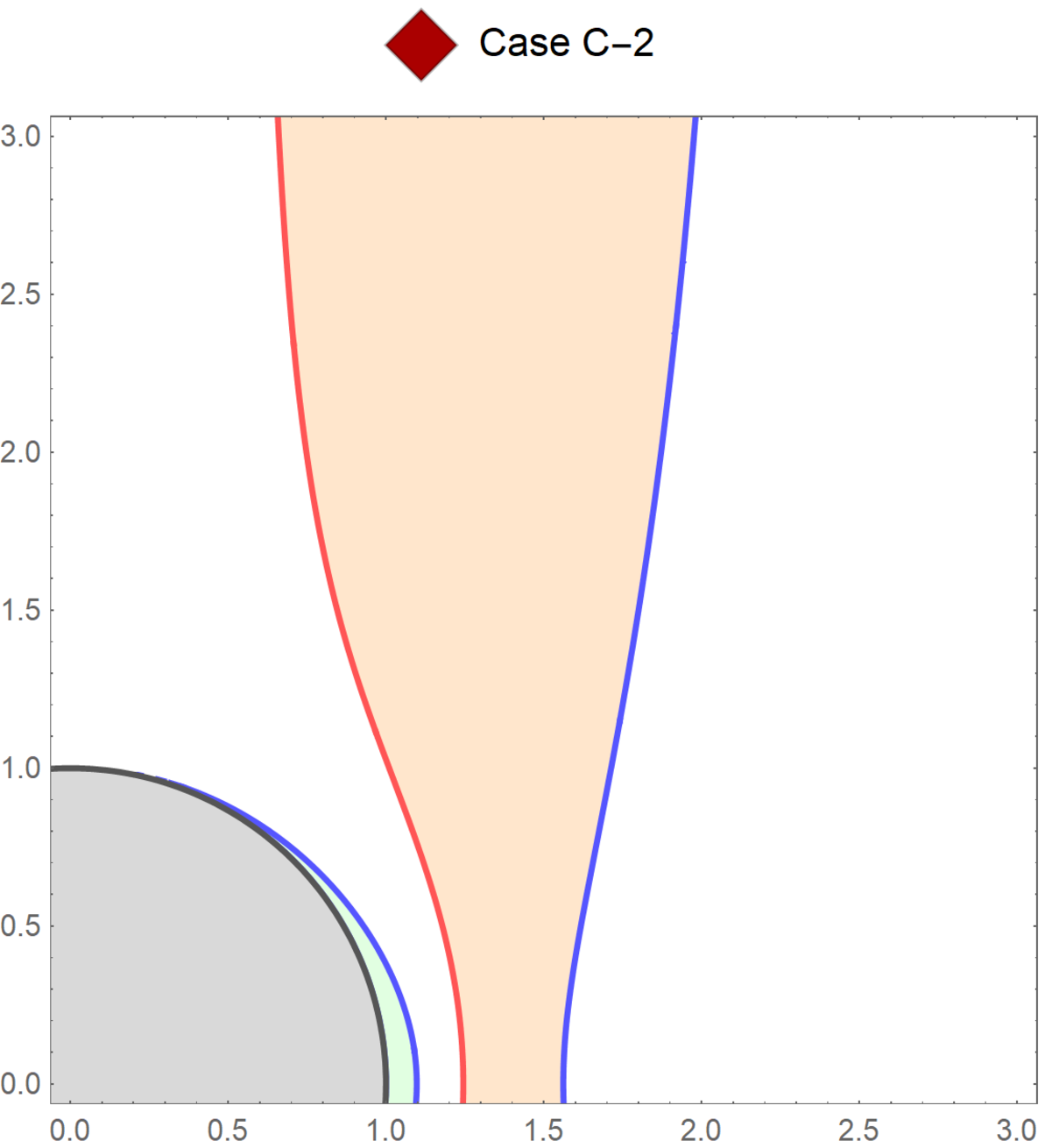}
\includegraphics[scale=0.3]{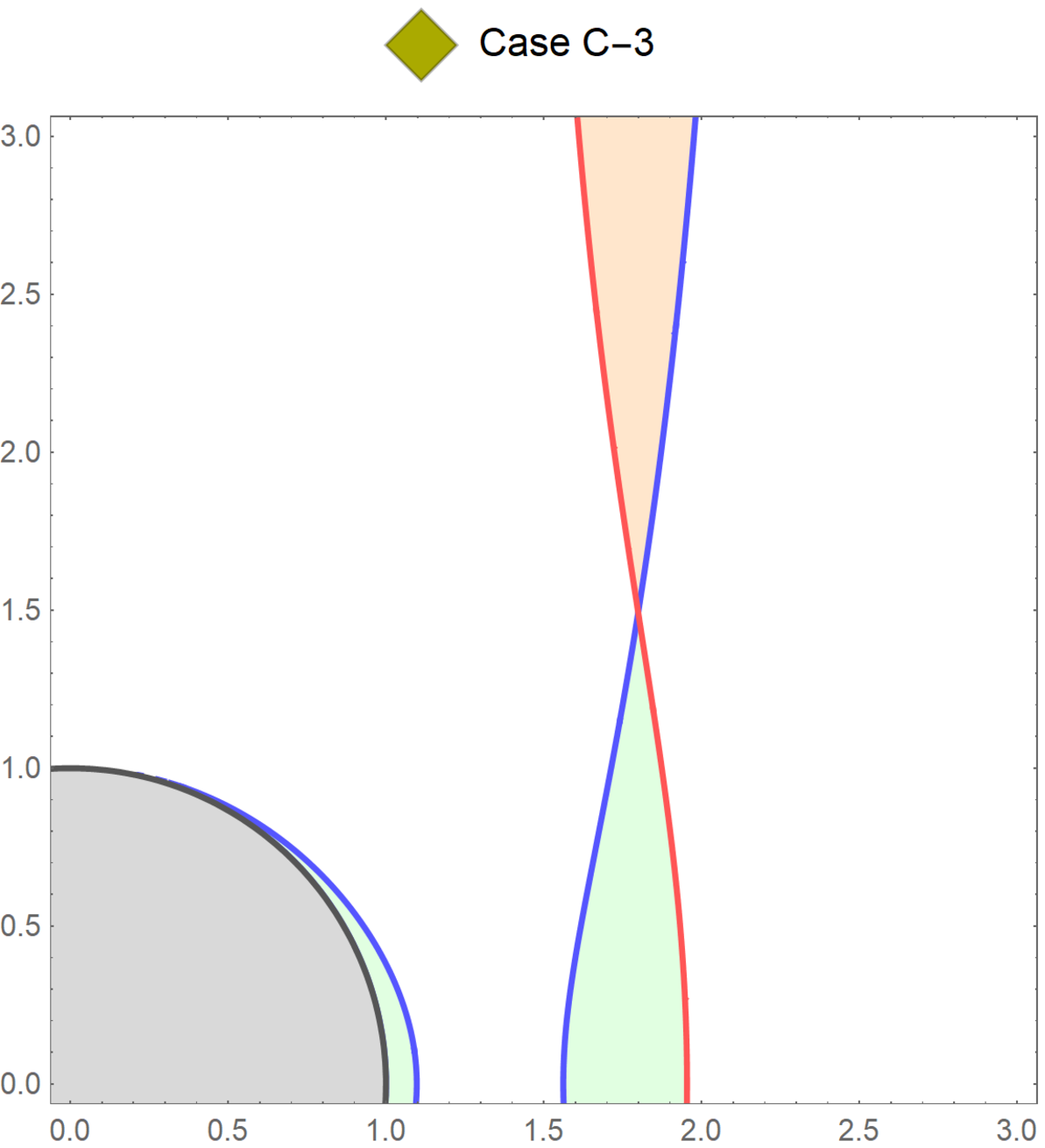}
\includegraphics[scale=0.3]{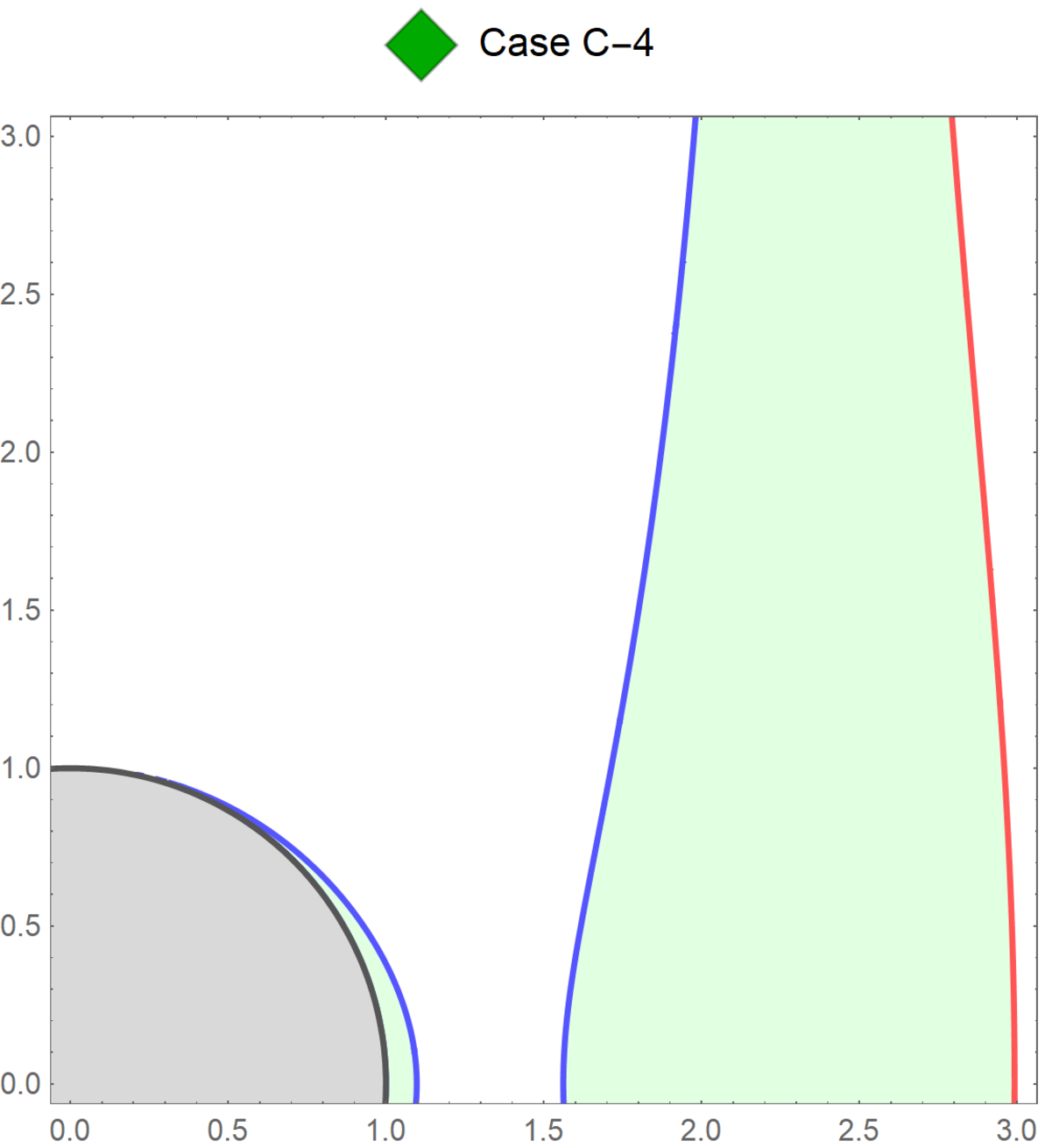}\\
\vspace{7mm}
\includegraphics[scale=0.3]{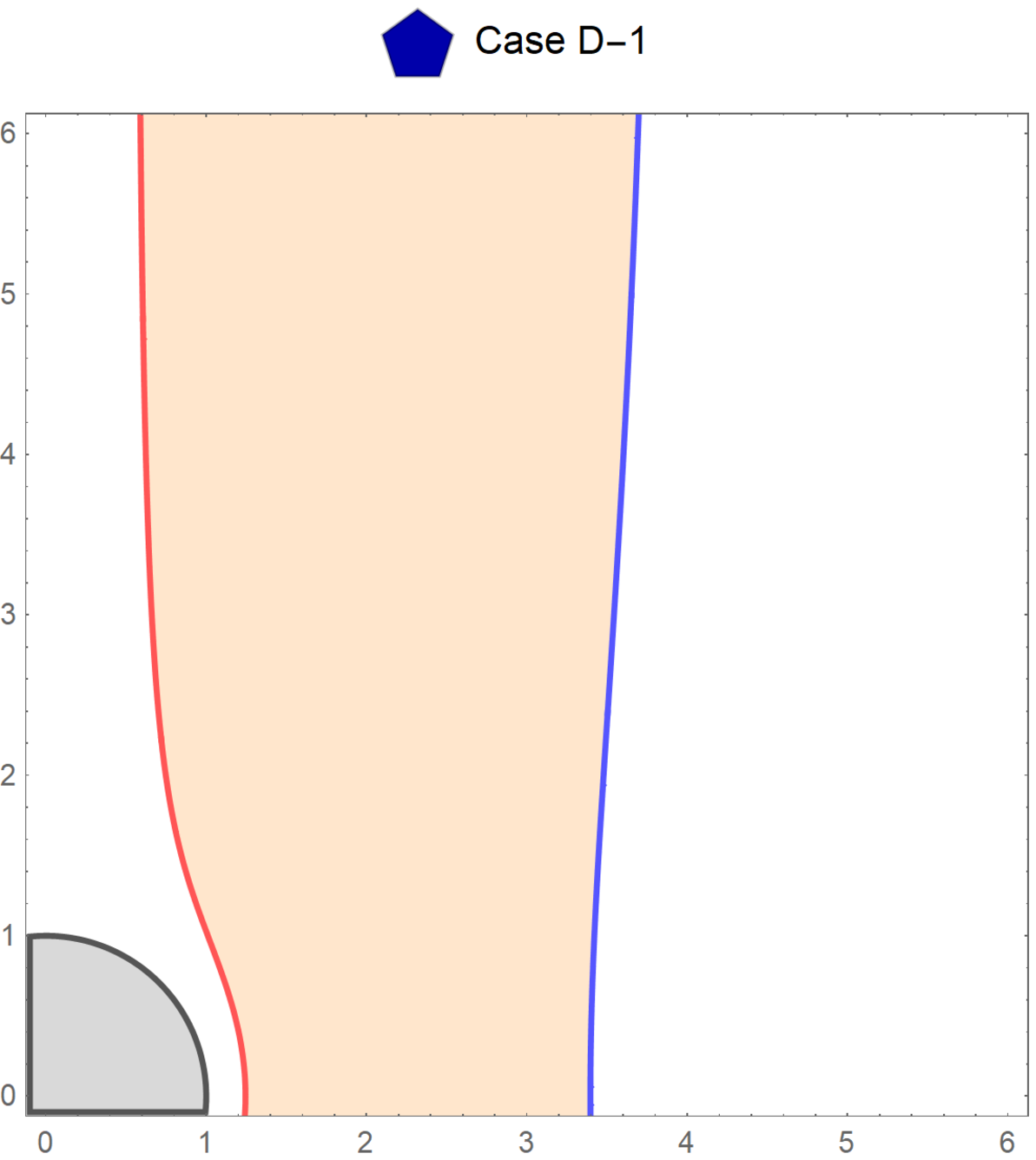}
\includegraphics[scale=0.3]{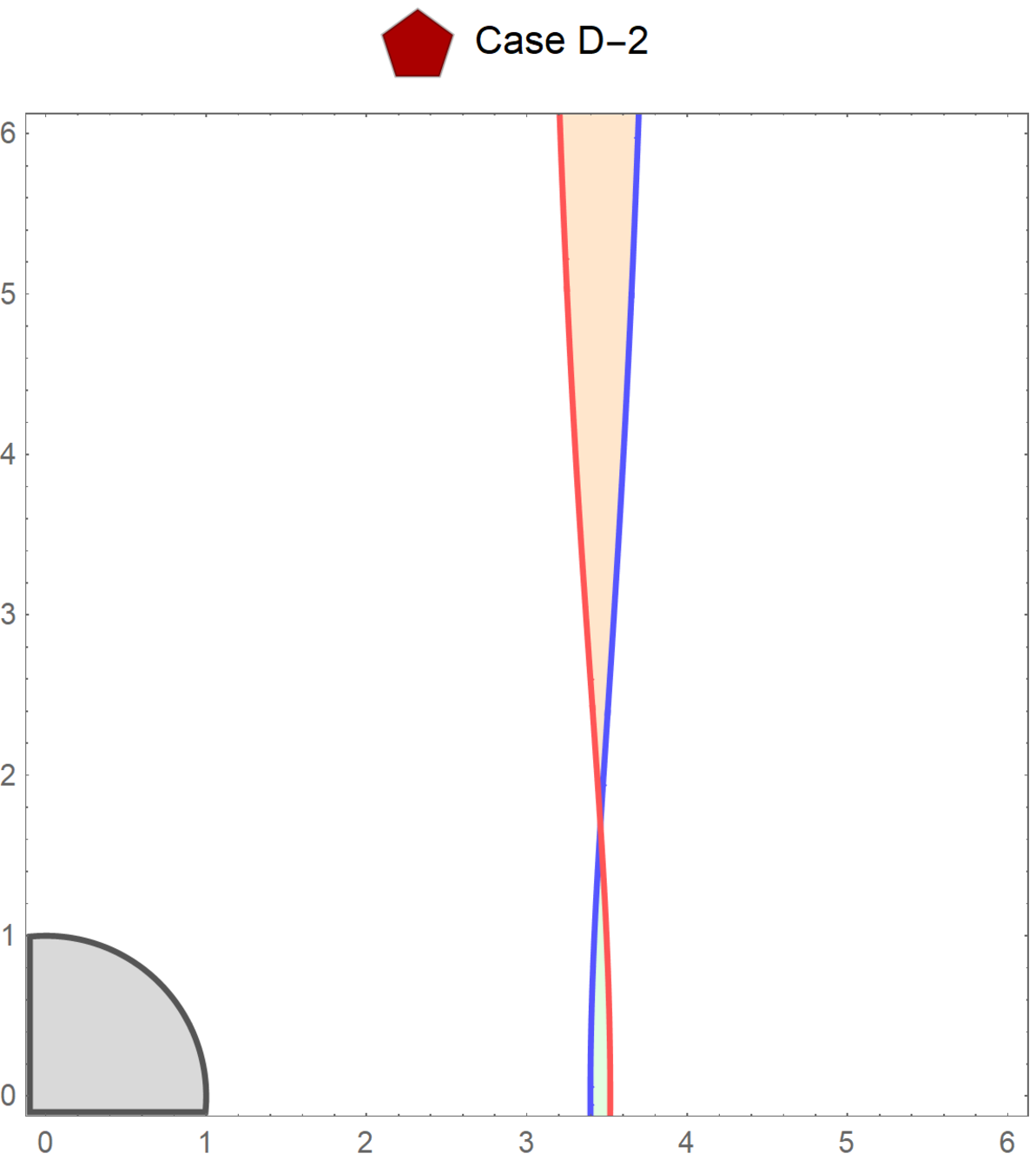}
\includegraphics[scale=0.3]{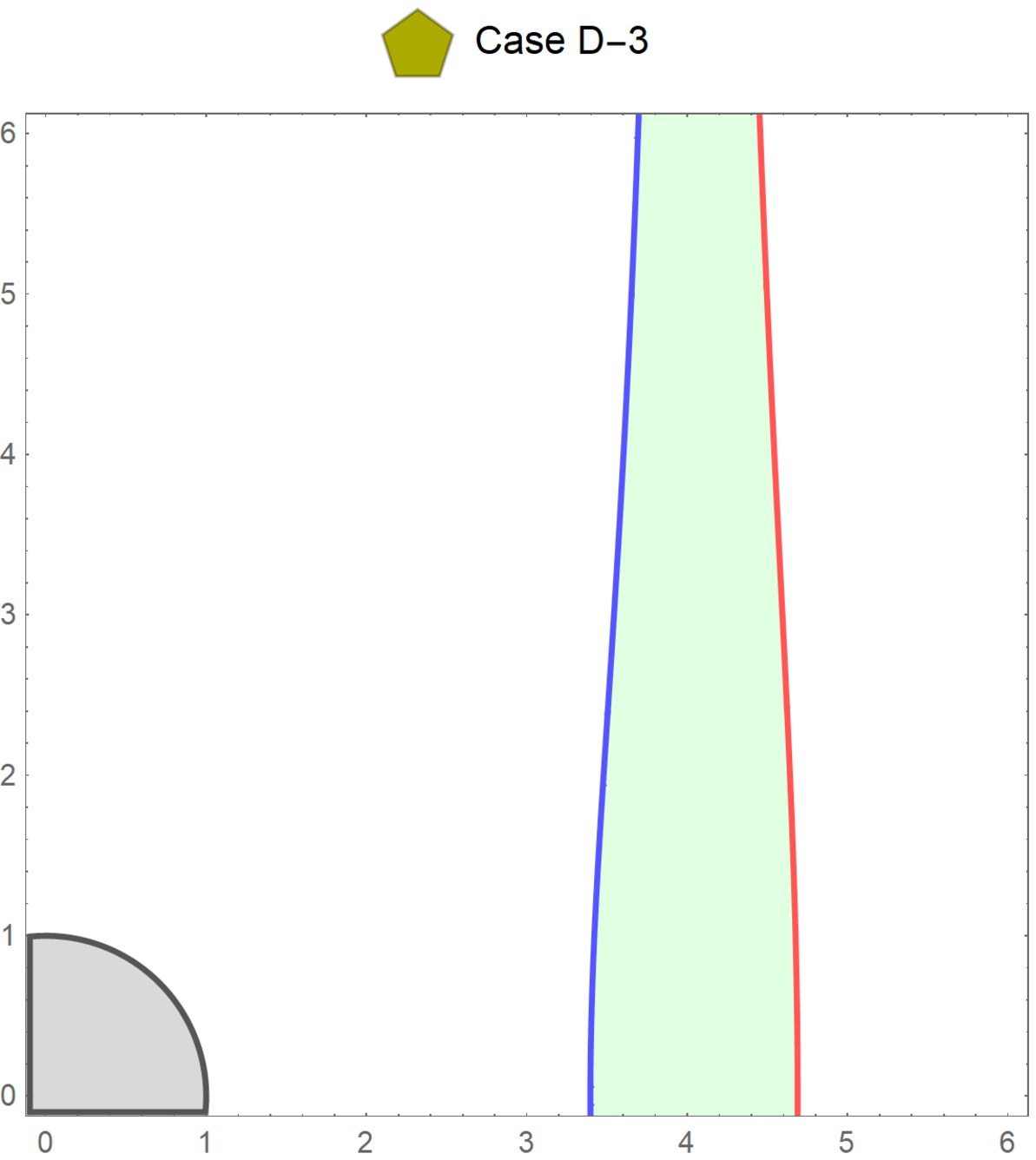}
\caption{\baselineskip5.5mm
The allowed region for the parameter values 
$q$ and $\Omega$ with $a=M/2$ marked in Fig.~\ref{fig:pararegion}
are depicted. 
The length scale is 
normalized by $r_+$. 
}
\label{fig:cases}
\end{center}
\end{figure}

%

The intersecting point S is a saddle point of the potential $V$ in the $r$-$\theta$ plane. 
Stationary rotating strings exist in the region $f<0$ and $k>0$,
and non-stationary rotating strings exist in the region $f>0$ and $k<0$.
A rigidly rotating string passing through the point S from the outside to the 
inside is converted from 
a stationary rotating string to a non-stationary rotating string. 


\newpage

\subsection{Regularity on the saddle point}
\label{reginls}

Since $f > 0$ in the vicinity of the horizon, a string that is stationary rotating 
in the far region needs to pass through the point S 
and be converted to a non-stationary rotating string
for sticking into the horizon. 
We consider whether the string can pass through S regularly. 

At the point S, 
we find that $p_r$ and $p_\theta$ must vanish because of the constraint \eqref{constr}. 
Then, Eqs.~\eqref{eom_r} and \eqref{eom_th} reduce to
\begin{align}
	\dot r=0 \quad{\rm and}\quad
	\dot \theta=0
\end{align}
at the point S. 
It means that the parameter $\sigma$ is not appropriate to analyze the 
behavior of the strings passing through the point S.  

Introducing the new parameter $\chi$ defined by 
\begin{equation}
	\frac{\dd \sigma}{\dd \chi}=\frac{1}{p_r}, 
\label{new_parameter}
\end{equation}
we obtain the equations with respect to $\chi$ across the point S in the form
\begin{align}
	r'&=\Delta, 
\label{r'}
\\
	p'_r&=-\frac{1}{2}\frac{p'_r\Delta \del_r^2 V  
		+p'_\theta \del_r \del_\theta V }{{p'_r}^2}, 
\label{pr_prime}\\
	\theta'&=\frac{p'_\theta}{p'_r}, 
\label{theta'}
\\
	p'_\theta&=-\frac{1}{2}\frac{p'_r\Delta \del_r\del_\theta V 
		+p'_\theta \del_\theta^2 V}{{p'_r}^2}, 
\label{ptheta_prime}
\end{align}
where the prime \lq$~'$\rq\ denotes the derivative with respect to $\chi$. 
Since $\Delta>0$ outside the horizon, $r'>0$ means that 
the parameter $\chi$ increases from the black hole side through S to the far side. 
From Eqs.~\eqref{pr_prime} and \eqref{ptheta_prime}, 
we obtain the following expressions for $p'_\theta$ and $p'_r$:
\begin{align}
	p'_\theta&=\frac{-p'_r\Delta\del_r\del_\theta V}{\del_\theta^2 V+2{p'_r}^2}, 
\label{p_theta'}
\\
	{p'_r}^2&=\frac{1}{4}\left(-\del_\theta^2 V-\Delta \del_r^2 V
		+\sqrt{\left(\del_\theta^2 V-\Delta\del_r^2 V\right)^2
		+4\Delta\left(\del_r\del_\theta V\right)^2}\right). 
\label{p_r'}
\end{align}
Evaluating Eqs.~\eqref{p_theta'} and \eqref{p_r'} at S, we find 
$p'_r$ and $p'_\theta$ are finite at the point S, 
and $\theta'$ is also finite at the point S.  
We note that the sign of $p_r$ changes across the point S, then 
the sign of $f p_r$ does not change.

\subsection{Induced Metric}

The metric induced on the world sheet of the string is defined by
\begin{equation}
	\gamma_{AB}=g_{\mu\nu}
			\frac{\del x^\mu}{\del \zeta^A}
		\frac{\del x^\nu}{\del \zeta^B}, 
\label{eq:indmet}
\end{equation}
where $\zeta^A=(\tau,\sigma)$ are parameters on the world sheet. 
If we take $\sigma$  by Eq.~\eqref{laps} and  
$\tau=t$ being a parameter along the integral curve of $\xi$,  
using the equations of motion~\eqref{eom_r}, 
\eqref{eom_th}, and \eqref{eom_ph},  we have 
the induced metric in the form 
\begin{equation}
	\gamma_{AB}\dd \zeta^A\dd \zeta^B
		=f\dd \tau^2+\frac{2qf}{\Delta}\left(-2Mra+A\Omega\right)
	\dd \tau\dd\sigma
	+\left(-fk\Sigma^2+\frac{q^2f^2\Sigma A}{\Delta^2\sin^2\theta}\right)\dd\sigma^2.  
\end{equation}
We find that 
the determinant of the induced metric given by
\begin{equation}
	\det\gamma=-f^2\Sigma^2
\end{equation}
is negative except at the point S. 

We use $\chi$ introduced by Eq.~\eqref{new_parameter} 
instead of $\sigma$ on the world sheet near the point S, 
then the induced metric is obtained by using Eqs.~\eqref{r'} and \eqref{theta'} 
with Eq.~\eqref{p_theta'} in a regular form: $-f^2\Sigma^2/p_r^2$.  
The regularity on the light surface discussed in Sec.~\ref{reginls} 
guarantees that 
the world sheet is everywhere timelike.

\section{Regularity on the horizon}
\label{sec:reghori}

\begin{figure}[!htbp]
\begin{center}
\includegraphics[scale=0.6]{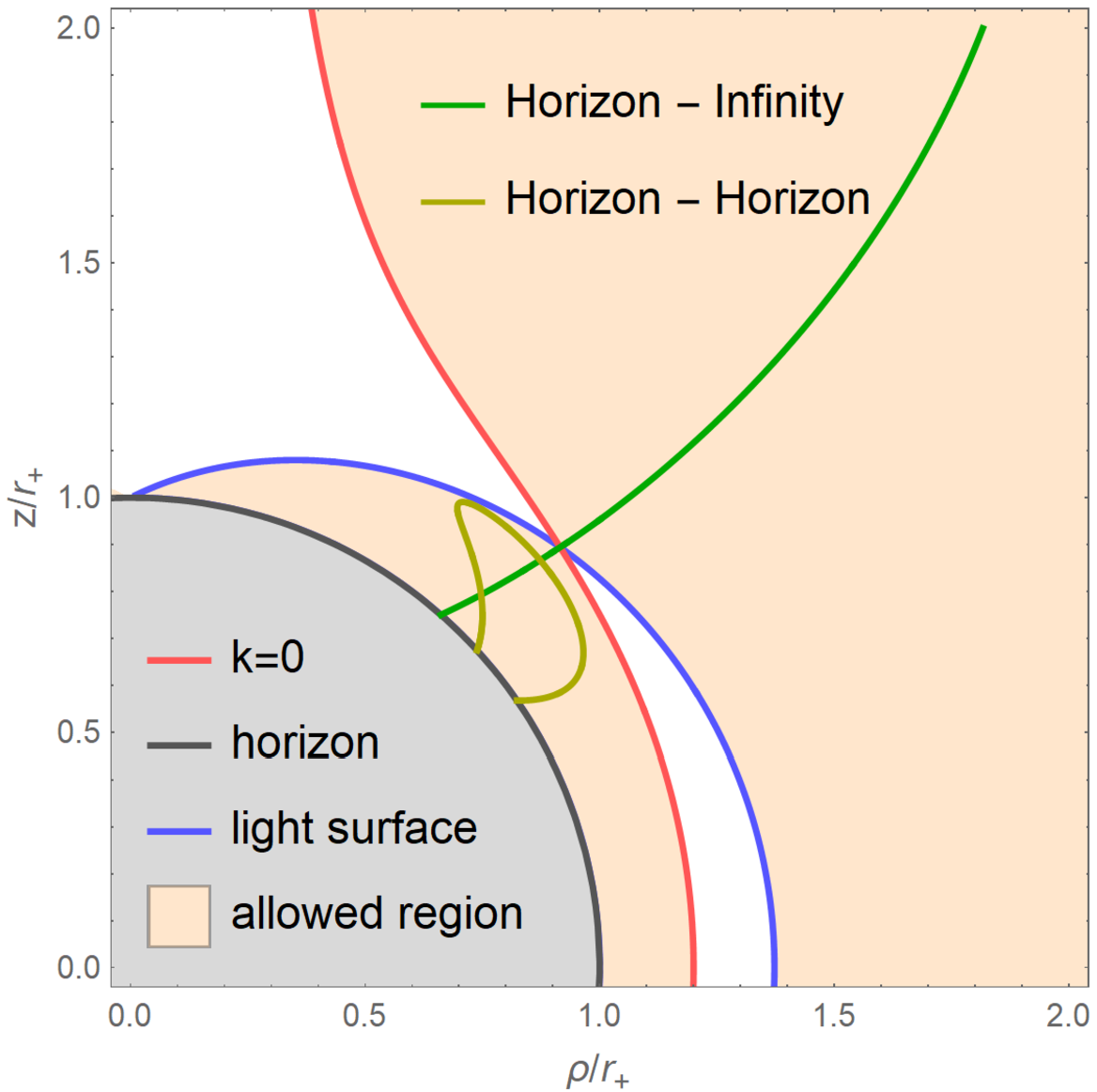}\hspace{1cm}
\includegraphics[scale=0.6]{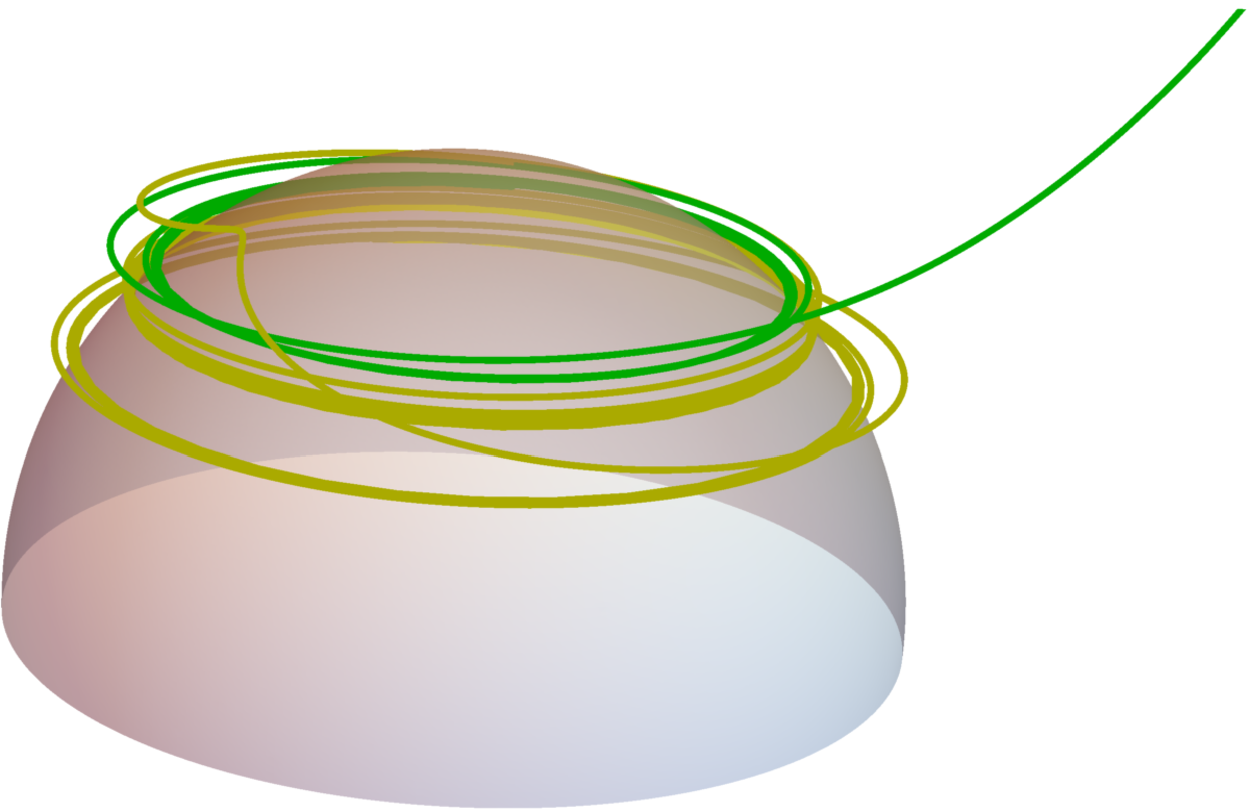}
\caption{\baselineskip5.5mm
Two cases of the string configuration projected on the $\rho$-$z$ plane(left) and those three dimensional view(right), 
where $(x,y,z)=(r\sin\theta\cos\phi, r\sin\theta\sin\phi, r\cos\theta)$ and $\rho:=r\sin\theta$(left). 
One is the string connecting a horizon neighbourhood and infinity, and the
other is string with both ends that exist in a horizon neighbourhood.
%
%
}
\label{fig:trajectoryBL}
\end{center}
\end{figure}

Let us show two cases of the string configuration on a constant $t$ slice 
in the Boyer--Lindquist coordinates. 
Solving Eqs.~\eqref{eom_r}--\eqref{eom_ph} with $a/M=1$, $\Omega M=1/4$, and $q/M=0.2$, 
we obtain two cases: a string which connects a black hole neighborhood and infinity, 
and a string whose both ends exist in a horizon neighborhood 
(see Fig.~\ref{fig:trajectoryBL}). 
These strings wind infinitely on the horizon 
because $\dot \phi$ diverges in the limit to the horizon as is seen in Eq.~\eqref{eom_ph}. 
Note that, however, 
such infinite winding occurs at the bifurcation two-sphere, 
and it does not necessarily indicate the prohibition of horizon penetration in general.

In order to investigate possible configurations of the rigidly rotating 
string near the horizon and the 
condition for sticking into the horizon, let us rewrite the 
equations of motion in the Kerr coordinate, which is regular on the horizon.
The line element is given by
\begin{align}
\dd s^2
	&=-\left(1-\frac{2Mr}{\Sigma}\right)\dd v_\pm^2\pm 2\dd v_\pm \dd r
	+\Sigma \dd \theta^2
\cr
	&+\frac{A}{\Sigma}\sin^2\theta\dd \bar \varphi_\pm^2
		\mp 2a\sin^2\theta\dd r\dd \bar \varphi_\pm 
	-\frac{4Mra}{\Sigma}\sin^2\theta\dd v_\pm\dd\bar \varphi_\pm, 
\end{align}
where the subscripts $+$ and $-$ denote the advanced and retarded 
Kerr coordinates, respectively. 

The Killing vector $\xi$ is given by 
$\del_{v_\pm}+\Omega\del_{\bar \varphi_\pm}$ in the present coordinate. 
By the same procedure as was used in Sec.~\ref{sec:eom}, 
we obtain the reduced space metric in the form
\begin{align}
	\tilde h_{ij}\dd x^i \dd x^j
=\left(1-a\Omega\sin^2\theta\right)^2\dd r^2-f\Sigma \dd \theta^2
		+\Delta \sin^2\theta \dd \varphi_\pm^2
		+2C_\pm \sin^2\theta \dd r\dd \varphi_\pm, 
\label{ind_met_Kerr}
\end{align}
where $\varphi_\pm=\bar \varphi_\pm -\Omega v_\pm$, 
and we have defined $C_\pm$ as 
\begin{align}
	C_\pm &:=\pm\left[\Omega\left(r^2+a^2\right)-a\right]. 
\end{align}
Since the reduced metric \eqref{ind_met_Kerr} has the same form for the advanced 
and retarded Kerr coordinates, 
hereafter, 
we concentrate on the future event horizon and drop the subscript $+$ 
for notational simplicity.  
Then, the Hamiltonian is given by 
\begin{align}
	{\cal H}
	=-\frac{N}{2f\Sigma}\left(\Delta p_r^2+p_\theta^2-2C p_r p_\varphi 
	+\frac{\left(1-a\Omega\sin^2\theta\right)^2}{\sin^2\theta}p_\varphi^2\right)
		-\frac{N}{2}, 
\end{align}
where, since $\partial_\phi=\partial_\varphi$, we find $p_\phi=p_\varphi=-q$. 
The constraint equation is given by 
\begin{equation}
(\Delta p_r+Cq)^2+\Delta p_\theta^2
	-f\Sigma\left(\frac{q^2}{\sin^2\theta}-\Delta\right)=0.  
\end{equation}
Setting $N=-f\Sigma$, 
we obtain the following equations of motion 
\begin{align}
	\dot r&=\Delta p_r +q C,
\label{eq:dr}
\\
	\dot p_r&= 	(M-r)p_r^2  -q  p_r \del_rC
  		-\frac{1}{2}\del_r\left(f\Sigma\right), 
\\
	\dot \theta&=p_\theta, \\
	\dot p_\theta&=
		-\frac{q^2}{2}\del_\theta\left(\frac{(1-a\Omega\sin^2\theta)^2}{\sin^2\theta}\right)
		-\frac{1}{2}\del_\theta\left(f\Sigma\right),\\
	\dot \varphi &=-\frac{q(1-a\Omega\sin^2\theta)^2}{\sin^2\theta}-Cp_r, 
\label{eq:dvphi}
\end{align}
where the dot \lq$\dot ~$\rq\ denotes the derivative with respect to 
the parameter along the geodesic on the metric \eqref{ind_met_Kerr}. 

It is different from the previous equations (\ref{eom_r})--(\ref{eom_ph})
with the Boyer--Lindquist coordinates 
that these equations can describe regular evolution of a string on the horizon. 
If we set $\dot r <0$ on the horizon, namely such a string is sticking into the horizon, 
where $\Delta=0$ in Eq.~\eqref{eq:dr}, then  $qC$ should be negative. 
It means that the following inequality holds: 
\begin{equation}
	q\left(\Omega_{\rm H}-\Omega\right)\geq 0 . 
\label{eq:increasingarea}
\end{equation}
Since $\dot r <0$ and $f>0$ on the horizon, 
then the quantity $q$ is regarded as the outward angular momentum flux 
(see Appendix~\ref{sec:flux}). 
This inequality means the non-negativity of 
the energy flux across the horizon, 
which is measured by the stationary null observer on the horizon 
(i.e., the horizon generator $\partial_v+\Omega_{\textrm{H}} \partial_{\bar{\varphi}}$).

\begin{figure}[htbp]
\begin{center}
\includegraphics[scale=0.5]{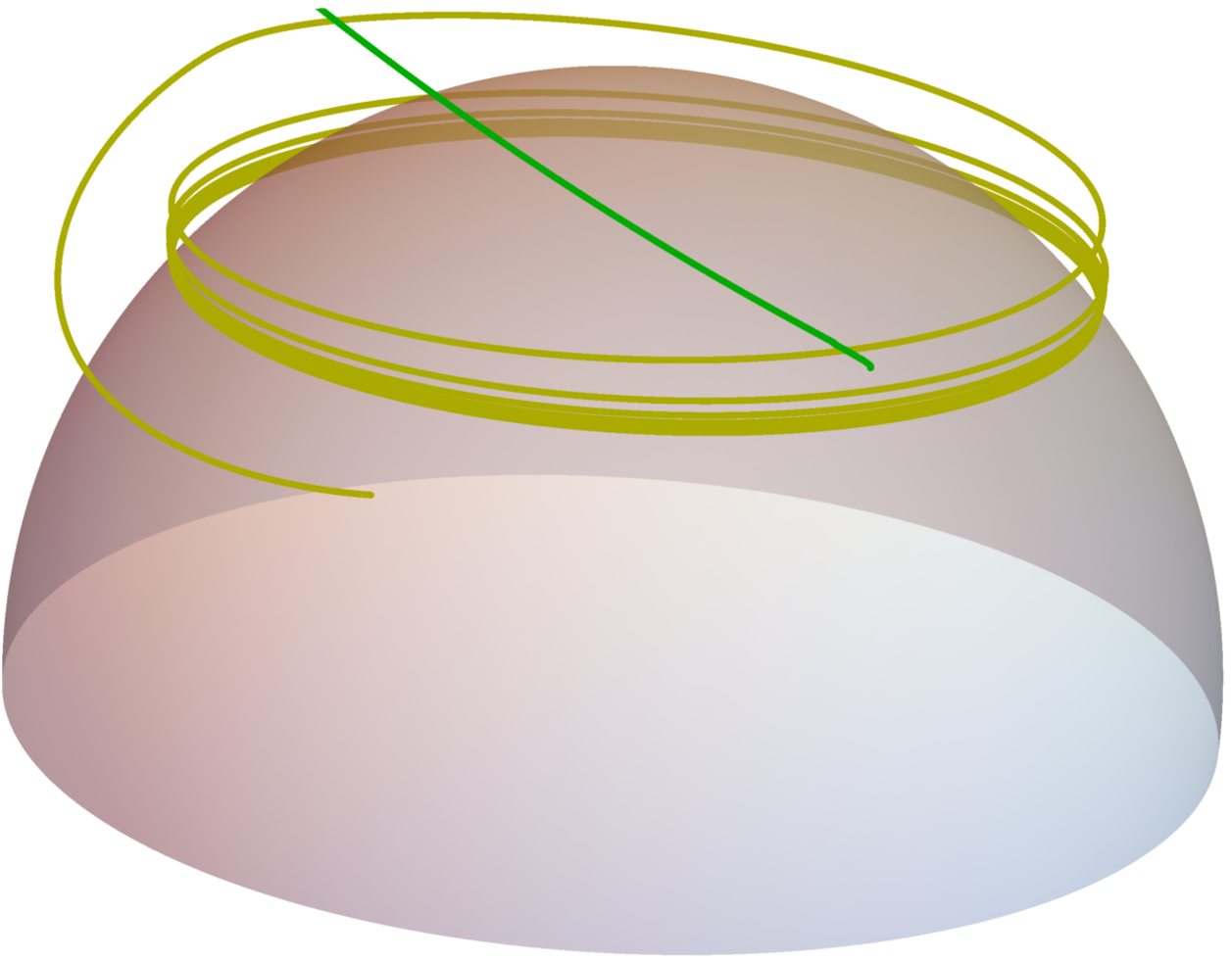}
\includegraphics[scale=0.6]{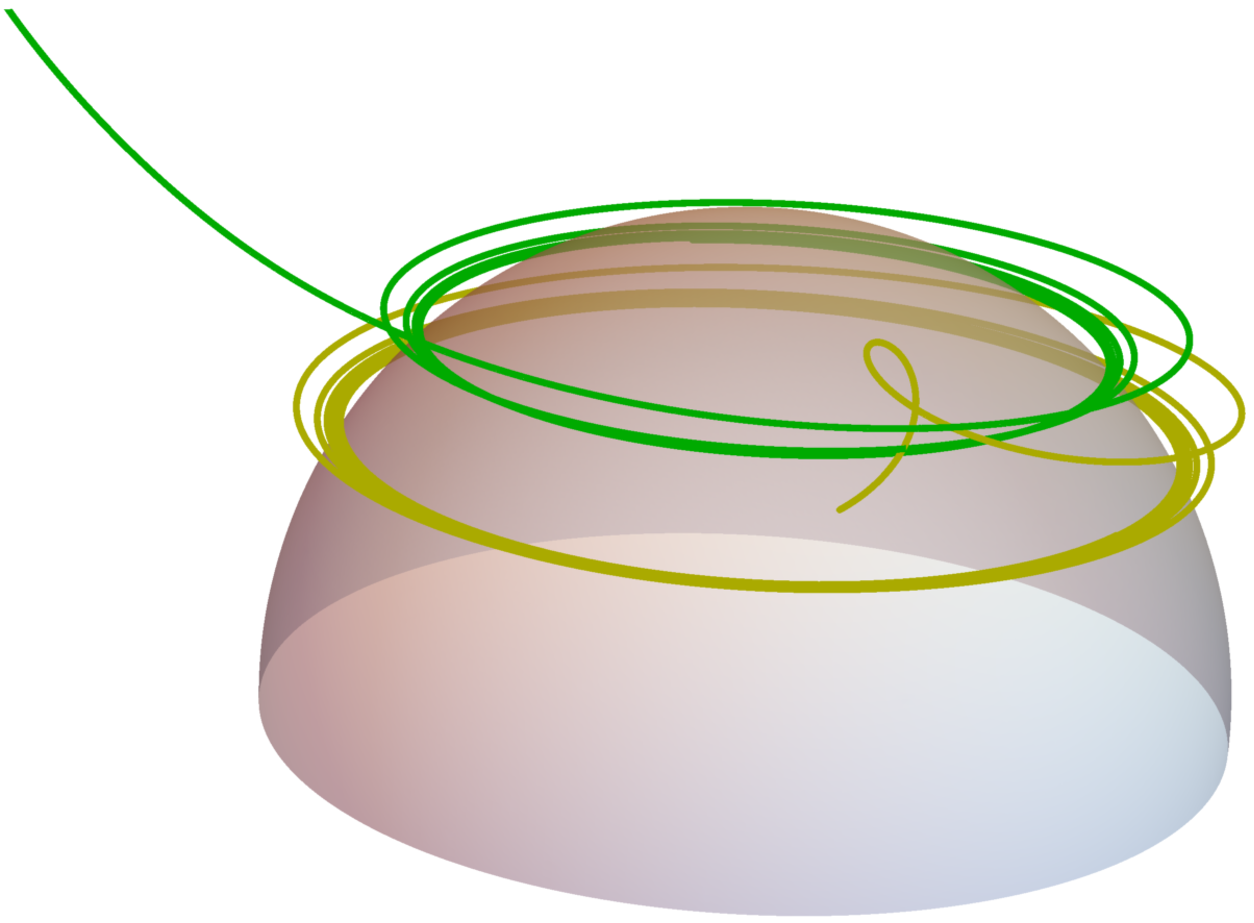}
\caption{\baselineskip5.5mm
Three dimensional view of the two cases of the string configuration, 
in the advanced Kerr coordinates(left) and 
retarded Kerr coordinates(right), 
where $(x,y,z)=(r\sin\theta\cos\varphi_\pm, r\sin\theta\sin\varphi_\pm, r\cos\theta)$. 
One is the string connecting a horizon neighbourhood and infinity, and the
other is string with both ends that exist in a horizon neighbourhood.
}
\label{fig:trajectory3D}
\end{center}
\end{figure}

Conversely, any rigidly rotating string that does not 
satisfy the condition \eqref{eq:increasingarea} 
cannot penetrate the horizon. 
As is shown in Fig.~\ref{fig:trajectory3D}, there are two possible behaviors 
near the horizon, i.e.,
the sticking into the horizon or infinitely winding around the horizon. 
It should be pointed out that both ends of a rigidly rotating string 
cannot stick into the horizon. Since $q$ is a constant that describes 
the constant flow of the angular momentum along the string, noting that the sign of $\dot r$ changes at the other end, 
we find that only one of the end can satisfy the condition \eqref{eq:increasingarea}. 
By using the retarded Kerr coordinates, 
we obtain the condition for sticking into the past event horizon as 
\begin{equation}
	q\left(\Omega_{\rm H}-\Omega\right)\leq 0 . 
\label{eq:past}
\end{equation}

\section{Energy extraction from the black hole}
\label{sec:maxlumi}

We consider the energy extraction from the Kerr black hole by using a  
rigidly rotating string 
which sticks into the black hole~\cite{Kinoshita:2016lqd}. 
We fix the direction of the parameter $\sigma$ as $\dot r <0$ near the horizon. 

Within the parameter range 
of a totally non-stationary rotating string, for example 
the case of A-2 in Fig.~\ref{fig:cases}, 
the condition for the outward energy flux, $q\Omega >0$, and the condition 
for sticking into the horizon, \eqref{eq:increasingarea}, are not compatible.

In the case of a stationary rotating string, on the other hand, as is noted above, 
a string needs to pass through the point S for sticking into the horizon. 
Since the sign of $f$ flips at S, then the sign of $\dot r$ does simultaneously. 
Therefore, $q$ and $q\Omega$ mean the outward flux of angular momentum and energy, 
respectively, throughout the string (see Appendix~\ref{sec:flux}). 
The energy extraction from the Kerr black hole by 
a stationary rotating string can be realized in the limited parameter region where $q\Omega >0$ and 
the condition \eqref{eq:increasingarea} holds(see Fig.~\ref{fig:pararegion}). 
The condition \eqref{eq:increasingarea} means the second law of 
the black hole thermodynamics: 
\begin{align}
	T \dd S &= \dd M - \Omega_H \dd J 
\cr
&= q(\Omega_H -\Omega ) \geq 0, 
\end{align}
namely the area law~\cite{Kinoshita:2016lqd}. 
Every rigidly rotating string passing through the horizon is 
compatible with the area 
law of the black hole. 

The maximum value of the energy flux  extracted 
from the black hole is realized for the case in which the curve $\ell:=q\Omega={\rm const.}$ 
is tangent to the curve \eqref{boundary_curve_2}. 
Then we have
\begin{equation}
	\ell=\frac{q(a-q)}{2M^2+2M\sqrt{M^2-a^2+q^2}+q^2-qa}. 
\label{eq:formaxp}
\end{equation}
Since the right-hand side is an increasing function of $a$ for $q>0$ and $\Omega>0$, 
we consider the extremal case, that is, $a=M$. 

Then Eq.~\eqref{eq:formaxp} can be rewritten as follows:
\begin{equation}
	\ell=\frac{q(M-q)}{2M^2+M q+q^2}. 
\label{eq:formaxp_ex}
\end{equation}
We can  easily  find that the maximum value is
\begin{equation}
	\ell_{\rm max}=\frac{4\sqrt{2}-5}{7}. 
\end{equation}
Therefore, the maximum power from the Kerr black hole through a test Nambu--Goto string is given by 
$(4\sqrt{2}-5)/7\times G\mu/c^4 \simeq9.38\times10^{-2}\times G\mu/c^4$ times the Dyson luminosity($c^5/G$), 
for $a=M$ and $\theta_{P+}=\pi/2$, where $\mu$ is the tension of the string and we have restored 
the $G$'s and $c$'s in this equation.

\section{Summary and Discussion}
\label{sec:sumdis}

We have analyzed the rigidly rotating co-homogeneity-1 string which is foliated by 
a stationary Killing vector field in the Kerr spacetime. 
The problem to obtain the string configuration is reduced to solving geodesic equations  
on a reduced three-dimensional space.  
We have 
shown that the world sheet is everywhere timelike. 
We have considered the regularity of the 
string configuration at two different positions, the light surface and the horizon. 
We have classified the configuration of the allowed region of the string motion
by three parameters per unit black hole mass;   
the string angular velocity, angular momentum flux and black hole spin. 
From the regularity on the horizon, 
we have found that 
a rigidly rotating string can penetrate the black(white) hole horizon. 
All such configurations have non-negative physical energy flux through the horizon, 
and are consistent with the area law of the black hole thermodynamics. 
If the string with outward(inward) physical energy flux 
stretches to the black(white) hole horizon, 
it 
is infinitely wound around the horizon. 
On the bifurcation two sphere, the string is infinitely wound as was reported in Ref.~\cite{Frolov:1996xw}. 
It has also been shown that the maximum value of the luminosity of the energy extraction 
from a Kerr black hole through a Nambu--Goto string is given by 
$(4\sqrt{2}-5)/7 \times \mu c$.

\section*{Acknowledgements}
We thank S.~Kinoshita for helpful comments. 
This work was supported by JSPS
KAKENHI Grant Numbers JP16K17688, JP16H01097 (CY), JP16K05358 (HI) 
and 
MEXT-Supported Program for the Strategic
  Research Foundation at Private Universities, 2014--2017, S1411024~(TI).

\appendix

\section{Number of roots of the equation $f(r)=0$}
\label{sec:Xr0}

On the restriction of the range $r>r_+$, the roots of $f=0$ are the same as $r f=0$, 
which gives a cubic equation. 
The positive root $r_*$ of the equation $(rf)'=0$ is given by
\begin{equation}
	r_*=\sqrt{\frac{1-a^2\Omega^2}{3\Omega^2}}. 
\end{equation}
The condition that the equation $r f=0$ has two positive roots is  
\begin{align}
	&r_*>r_+  , \quad {\rm and }\quad r_*f(r_*) < 0. 
\label{two_roots}
\end{align}
For the existence of real $r_*$, $a^2\Omega^2 < 1$ must hold. 
These inequalities are equivalent to 
\begin{align}
	&1+2a^2\Omega^2-6M^2\Omega^2 >6M\Omega^2 \sqrt{M^2-a^2}, 
\label{ieq1}
\\ 
	&\left(1+a\Omega\right)^3-27 M^2\Omega^2(1-a\Omega) > 0 .  
\label{ieq2}
\end{align}
In Eq.~\eqref{ieq1}, since 
the right-hand side is non-negative, the left-hand side must be positive. 
Therefore, we obtain
\begin{equation}
1+2a^2\Omega^2>6M^2\Omega^2. 
\label{ieq1-1}
\end{equation}
Taking the square of the inequality \eqref{ieq1}, we obtain
\begin{equation}
\frac{\left(1+2a^2\Omega^2\right)^2}{1-a^2\Omega^2}>12M^2\Omega^2. \label{ieq1-2}
\end{equation}
It is easy to find that, under the condition \eqref{ieq2}, 
the inequality \eqref{ieq1-2} is automatically satisfied, and  
the condition \eqref{ieq1-1} can be replaced by 
\begin{align}
	|a\Omega| < \frac12.  
\label{range_Omega}
\end{align}
Then, the parameter region for the existence of two roots is given by Eq.~\eqref{eq:tworoot}. 
The parameters $\Omega_{\rm cr}^\pm$ are given by the positive and negative roots of Eq.~\eqref{eq:forcr} 
in the range \eqref{range_Omega} 
as follows:
\begin{align}
	\Omega_{\rm cr}^+ 
		= 2p \cos\left(\frac13 \arccos\left(\frac{q}{2p}\right)+\frac{2\pi}{3}\right)
				-\frac{a^2-9M^2}{a(27M^2+a^2)},
\\
	\Omega_{\rm cr}^- 
		= 2p \cos\left(\frac13 \arccos\left(\frac{q}{2p}\right)+\frac{4\pi}{3}\right)
				-\frac{a^2-9M^2}{a(27M^2+a^2)},
\end{align}
where 
\begin{align}
	p := \frac{3M\sqrt{9M^2-5a^2}}{a(27M^2+a^2)}, \quad 
	q := \frac{6(a^4-36M^2a^2+27M^4)}{a(9M^2-5a^2)(27M^2+a^2)}. 
\end{align}

\section{Parameters for the intersection of 
the light surface and the centrifugal barrier:~a simple derivation}
\label{sec:irrex}
\newcommand{\rs}{r_{\rm S}}
\newcommand{\ts}{\theta_{\rm S}}

The intersecting point S of the light surface $f=0$ and the centrifugal barrier 
$k=0$ exists for limited regions in the $\Omega$-$q$ plane. 
The parameter regions for the existence of the intersecting 
point are bounded by two kinds of curves. 
The first kind curves specify that 
the intersecting point disappears at $r_{\rm S}\to\infty$ and $\theta_{\rm S}\to 0$. 
The curves are given by 
\begin{align}
	&r f \to -1+\Omega^2 \rs^2 \sin^2\ts =0,
\\
	&\Delta k \to 1- \frac{q^2}{\rs^2 \sin^2\ts}=0. 
\end{align}
Then, eliminating $\rs^2\sin^2\ts$ from these equations, we have 
\begin{align}
	\Omega^2 q^2 = 1.
\end{align}
The second kind curves specify that the intersecting point S disappears 
on the equatorial plane. 
Then, curves are determined by
\begin{align}
	&r f(\rs, \ts =\frac{\pi}{2}) 
		= \Omega^2 \rs^3-(1-a^2 \Omega^2) \rs +2M(1-a\Omega)^2  =0, 
\\
	&\Delta k(\rs, \ts =\frac{\pi}{2}) =\rs^2-2M\rs+a^2-q^2=0. 
\end{align}
Eliminating $\rs$ from these equations, we have
\begin{align}
	&\Omega^4 q^6 
	- \Omega ^2 \left(a^2 \Omega^2+2\right) q^4
\cr
	&- \left(a^2 
   \left(8 M^2 \Omega ^4-2 \Omega ^2\right)-24 a M^2 \Omega ^3
 	+16 M^2 \Omega ^2-1\right) q^2
\cr
	&-\left(4 a M^2 \Omega ^2+a-4 M^2 \Omega \right)^2 =0 . 
\end{align}
This equation determines the curves implicitly in the $\Omega$-$q$ plane 
once $M$ and $a$ are fixed. 

%

\section{Parameters for the intersection of 
the light surface and the centrifugal barrier:~details}
\label{sec:irrex_det}

Eliminating $\theta$ by using equations $f=0$ and 
$k=0$, 
we obtain the following equation:
\begin{equation}
P(r;q)P(r;-q)=0
\label{intseq}
\end{equation}
with
\begin{equation}
P(r;q):=r^2-\frac{2M}{1+q\Omega}r+a^2-aq, 
\end{equation}
where we have assumed $|q\Omega|\neq 1$. 
The case of $|q\Omega|\neq 1$ will be clarified below. 
Since the equation is symmetric under flipping the sign of $q$, 
hereafter, we consider only the case $q\geq0$ for a while. 

\subsubsection{Possible roots}
\label{sec:roots}


From $k=0$ we have the condition
\begin{equation}
0\leq\sin^2\theta_{\rm S}=\frac{q^2}{\Delta(r_{\rm S})}\leq1. 
\label{condth}
\end{equation}
The last inequality says
\begin{equation}
Q(r_{\rm S}):=r_{\rm S}^2-2Mr_{\rm S}+a^2-q^2\geq0. 
\label{ineq1}
\end{equation}
Let $r^{\rm ex}_{P\pm}$, $r_{P\pm}$, $r^{\rm ex}_Q$, and $r_Q$ 
denote roots for $P(r;\pm q)=0$ and $Q(r)=0$, respectively, as follows:
\begin{eqnarray}
r^{\rm ex}_{P\pm}&=&\frac{M}{1\pm q\Omega}
-\sqrt{\left(\frac{M}{1\pm q\Omega}\right)^2-a^2\pm q a}, 
\\
r_{P\pm}&=&\frac{M}{1\pm q\Omega}
+\sqrt{\left(\frac{M}{1\pm q\Omega}\right)^2-a^2\pm q a}, 
\\
r^{\rm ex}_Q&=&M-\sqrt{M^2-a^2+q^2}, 
\\
r_Q&=&M+\sqrt{M^2-a^2+q^2}. 
\end{eqnarray}
Since $r^{\rm ex}_Q<r_+$, the root $r^{\rm ex}_Q$ is irrelevant to the analysis.  
The inequality \eqref{ineq1} and $r^{\rm ex}_Q<r_+$ imply that $r_{P\pm}$ must satisfy $r_{P\pm}\geq r_Q$. 
As is shown in 
Ref.~\cite{Kinoshita:2016lqd}, 
$r^{\rm ex}_{P\pm}$ is irrelevant in the sense that, 
if $r^{\rm ex}_{P\pm}>0$, then $r^{\rm ex}_{P\pm}<r_Q$.
Therefore, only $r_{P\pm}$ are candidates for $r_{\rm S}$. 
In the case of $q\Omega=\pm1$, 
we obtain the equation $P(r;\pm1/\Omega)=0$ instead of Eq.~\eqref{intseq}.
Therefore we may just ignore $r_{P\mp}$ for $q\Omega=\pm1$.

\subsubsection{Region of $\Omega$ for the existence of the relevant root}

Regarding $r_P$ as a function of $\Omega$ and differentiating the equation $P(r_{P\pm};\pm q)=0$ by $\Omega$, 
we obtain
\begin{equation}
\pm\frac{\del r_{P\pm}}{\del \Omega}=\frac{-qMr_{P\pm}}{(1\pm q\Omega)^2\sqrt{[M/(1\pm q\Omega)]^2-a^2\pm qa}}\leq 0. 
\end{equation}
Then, we can find the following 
monotonic dependence:
\begin{equation}
\pm\frac{\del }{\del \Omega}\left(q^2/\Delta(r_{P\pm})\right)=\pm\frac{\del }{\del \Omega}\sin^2\theta_{P\pm}\geq0. 
\end{equation}
Here we note that the minimum value of $r_{P\pm}$ is given by $r_Q\geq r_+$ for $\theta_{P\pm}=\pi/2$. 
Therefore, for the existence of the relevant roots, the region of $\Omega$ is restricted as 
\begin{eqnarray}
&&\Omega_{0+}<\Omega\leq\Omega_{\frac{\pi}{2}+}~{\rm for~+branch}, \\
&&\Omega_{\frac{\pi}{2}-}\leq\Omega<\Omega_{0-}~{\rm for~-branch}, 
\end{eqnarray}
where $\Omega_{0\pm}=\mp1/q$ and 
$\Omega_{\frac{\pi}{2}\pm}$ is given by solving $Q(r_{P\pm})=0$ for $\Omega$ as 
\begin{equation}
\Omega_{\frac{\pi}{2}\pm}=\frac{a\mp q}{2M^2+2M\sqrt{M^2-a^2+q^2}+q^2\mp qa}. 
\end{equation}
The value of $\Omega_{0\pm}$ is given by taking 
the limit $r_{P\pm}\rightarrow \infty$ corresponding to $\sin\theta_{P\pm}=0$. 
Since we find 
\begin{equation}
\Omega_{\frac{\pi}{2}-}-\Omega_{\frac{\pi}{2}+}=\frac{2qr_Q}{2M(r_Q^2+a^2)+q^2r_Q}\geq0, 
\end{equation}
the two regions 
have intersection at $q=0$ only if $\Omega=\Omega_{\rm H}$. 
It is easy to show that the lines specified by $\Omega=\Omega_{\pm\frac{\pi}{2}}$ 
are equivalent to the 
lines given by 
Eq.~\eqref{boundary_curve_2}.

\subsubsection{Extremum of $\Omega_{\frac{\pi}{2}\pm}$ as a function of $q$}

In order to clarify the shape of the region which allows the existence of the intersection in the parameter space 
of $q$ and $\Omega$, let us regard $\Omega_{\frac{\pi}{2}\pm}$ as a function of $q$. 
The equation for the extremum $\dd\Omega_{\frac{\pi}{2}\pm}/\dd q=0$ can be written as 
\begin{equation}
2M^2-(a\mp q)^2=-\frac{2M\left[M^2-a(a\mp q)\right]}{\sqrt{M^2-(a^2-q^2)}}. 
\label{eq:qpm}
\end{equation}
For +branch(upper sign), taking the square of this equation, we obtain the following equation:
\begin{equation}
q^3-3aq^2-3(M^2-a^2)q+a(M^2-a^2)=0. 
\end{equation}
This equation has three roots $-q_-$, $q_0$, and $q_+$ satisfying 
$-q_-\leq0\leq q_0<q_+$ and $a<q_+$ for $a\leq M$, where the equality is given for $a=M$. 
It can be shown that each root of Eq.~\eqref{eq:qpm} is given by $q_\pm$(note we have assumed $q\geq0$). 
We can also show $\Omega_+:=\Omega_{\frac{\pi}{2}+}|_{q=q_+}<0$ 
from the following fact
\begin{equation}
\left.\Omega_{\frac{\pi}{2}+}\right|_{q=a}=0 ~{\rm and}
~\left.\frac{\dd\Omega_{\frac{\pi}{2}+}}{\dd q}\right|_{q=a}<0. 
\end{equation}
While, from $\Omega_-=\Omega_{\rm H}$ for $a=M$ and 
\begin{equation}
\left.\Omega_{\frac{\pi}{2}-}\right|_{q=0}=\Omega_{\rm H} ~{\rm and}
~\left.\frac{\dd\Omega_{\frac{\pi}{2}-}}{\dd q}\right|_{q=0}>0 ~{\rm for}~a<M,  
\end{equation}
we find $\Omega_-:=\Omega_{\frac{\pi}{2}-}|_{q=q_-}\geq\Omega_{\rm H}$.

\section{Energy and angular momentum flux}
\label{sec:flux}

Let us define the energy current $E^A$ on the string world sheet. 
We use the intrinsic coordinates 
$\zeta^A=(\tau,\sigma)$. 
The energy current is defined by 
\begin{equation}
	E^A:=-T^{AB}\frac{\del x^\mu}{\del \zeta^B}(\del_t)^\nu g_{\mu\nu}, 
\end{equation}
where $T^{AB}:=-\mu \gamma^{AB}$ is the string stress-energy tensor projected 
on the world sheet. 
The string tension $\mu$ is set to be unity hereafter for simplicity. 
Since $\del_t$ is the Killing vector field, $E^A$ is conserved, that is, 
\begin{equation}
	\mathcal D_A E^A 
	= \frac{1}{\sqrt{-\gamma}}\del_\tau\left(\sqrt{-\gamma}E^\tau\right)
		+\frac{1}{\sqrt{-\gamma}}\del_\sigma\left(\sqrt{-\gamma}E^\sigma\right)
	=0, 
\end{equation}
where $\mathcal D_A$ is the covariant derivative 
associated with the induced metric $\gamma_{AB}$. 
Then, the energy flux in the direction of increasing $\sigma$ 
is explicitly given by 
\begin{align}
	\sqrt{-\gamma}E^\sigma 
	&= \sqrt{-\gamma}\left(\gamma^{\sigma\tau}(g_{tt}+\Omega g_{\bar\phi t})
		+\gamma^{\sigma\sigma} \dot \phi g_{\bar\phi t}\right)
\cr
	&=-{\rm sign}(f)q\Omega. 
\end{align}
In the same way, 
we can define the angular momentum current by
\begin{equation}
	J^A
		:=T^{A B}\frac{\del x^\mu}{\del \zeta^B}(\del_\phi)^\nu g_{\mu\nu}. 
\end{equation}
The angular momentum flux in the direction of increasing $\sigma$ is given by 
\begin{align}
	\sqrt{-\gamma}J^\sigma 
	&= \sqrt{-\gamma}\left(\gamma^{\sigma\tau}(g_{t\bar\phi}+\Omega g_{\bar\phi\bar\phi})
		+\gamma^{\sigma\sigma} \dot \phi g_{\bar\phi\bar\phi}\right)
\cr
	&=-{\rm sign}(f)q. 
\end{align}


\begin{thebibliography}{10}
\baselineskip6mm
\bibitem{Lawrence:1993sg}
A.~E. Lawrence and E.~J. Martinec,
\newblock Phys. Rev. {\bf D50}, 2680 (1994), arXiv:hep-th/9312127, {\em {Black
  hole evaporation along macroscopic strings}}.

\bibitem{Rey:1998ik}
S.-J. Rey and J.-T. Yee,
\newblock Eur. Phys. J. {\bf C22}, 379 (2001), arXiv:hep-th/9803001, {\em
  {Macroscopic strings as heavy quarks in large N gauge theory and anti-de
  Sitter supergravity}}.

\bibitem{Maldacena:1998im}
J.~M. Maldacena,
\newblock Phys. Rev. Lett. {\bf 80}, 4859 (1998), arXiv:hep-th/9803002, {\em
  {Wilson loops in large N field theories}}.

\bibitem{Frolov:1988zn}
V.~P. Frolov, V.~Skarzhinsky, A.~Zelnikov, and O.~Heinrich,
\newblock Phys. Lett. {\bf B224}, 255 (1989), {\em {Equilibrium Configurations
  of a Cosmic String Near a Rotating Black Hole}}.

\bibitem{Vilenkin:2000jqa}
A.~Vilenkin and E.~P.~S. Shellard,
\newblock {\em {Cosmic Strings and Other Topological Defects}} (Cambridge
  University Press, 2000).

\bibitem{Ishihara:2005nu}
H.~Ishihara and H.~Kozaki,
\newblock Phys. Rev. {\bf D72}, 061701 (2005), arXiv:gr-qc/0506018, {\em
  {Classification of cohomogeneity one strings}}.

\bibitem{Koike:2008fs}
T.~Koike, H.~Kozaki, and H.~Ishihara,
\newblock Phys. Rev. {\bf D77}, 125003 (2008), arXiv:0804.0084[gr-qc], {\em {Strings
  in five-dimensional anti-de Sitter space with a symmetry}}.

\bibitem{Kozaki:2009jj}
H.~Kozaki, T.~Koike, and H.~Ishihara,
\newblock Class. Quant. Grav. {\bf 27}, 105006 (2010), arXiv:0907.2273[gr-qc], {\em
  {Exactly solvable strings in the Minkowski spacetime}}.

\bibitem{Morisawa:2017lpj}
Y.~Morisawa, S.~Hasegawa, T.~Koike, and H.~Ishihara,
\newblock (2017), arXiv:1709.07659[hep-th], {\em {Cohomogeneity-one-string
  integrability of spacetimes}}.

\bibitem{Frolov:1996xw}
V.~P. Frolov, S.~Hendy, and J.~P. De~Villiers,
\newblock Class. Quant. Grav. {\bf 14}, 1099 (1997), arXiv:hep-th/9612199, {\em
  {Rigidly rotating strings in stationary axisymmetric space-times}}.

\bibitem{Ogawa:2008qn}
K.~Ogawa, H.~Ishihara, H.~Kozaki, H.~Nakano, and S.~Saito,
\newblock Phys. Rev. {\bf D78}, 023525 (2008), arXiv:0803.4072[gr-qc], {\em
  {Stationary Rotating Strings as Relativistic Particle Mechanics}}.

\bibitem{Frolov:1995vp}
V.~P. Frolov, S.~Hendy, and A.~L. Larsen,
\newblock Phys. Rev. {\bf D54}, 5093 (1996), arXiv:hep-th/9510231, {\em {How to
  create a 2-D black hole}}.

\bibitem{Kinoshita:2016lqd}
S.~Kinoshita, T.~Igata, and K.~Tanabe,
\newblock Phys. Rev. {\bf D94}, 124039 (2016), arXiv:1610.08006[gr-qc], {\em {Energy
  extraction from Kerr black holes by rigidly rotating strings}}.

\bibitem{Ogawa:2008yx}
K.~Ogawa, H.~Ishihara, H.~Kozaki, and H.~Nakano,
\newblock Phys. Rev. {\bf D79}, 063501 (2009), arXiv:0811.2846[gr-qc], {\em
  {Perturbations of Spacetime around a Stationary Rotating Cosmic String}}.

\bibitem{Kozaki:2014aaa}
H.~Kozaki, T.~Koike, and H.~Ishihara,
\newblock Phys. Rev. {\bf D91}, 025007 (2015), arXiv:1410.6580[gr-qc], {\em {Membranes
  with a symmetry of cohomogeneity one}}.

\bibitem{Kinoshita:2017mio}
S.~Kinoshita and T.~Igata,
\newblock PTEP {\bf 2018}, 033E02 (2018), arXiv:1710.09152[gr-qc], {\em {The essence
  of the Blandford-Znajek process}}.

\bibitem{1971JETPL..14..180Z}
Y.~B. {Zel'Dovich},
\newblock Soviet Journal of Experimental and Theoretical Physics Letters {\bf
  14}, 180 (1971), {\em {Generation of Waves by a Rotating Body}}.

\bibitem{1972JETP...35.1085Z}
Y.~B. {Zel'Dovich},
\newblock Soviet Journal of Experimental and Theoretical Physics {\bf 35}, 1085
  (1972), {\em {Amplification of Cylindrical Electromagnetic Waves Reflected
  from a Rotating Body}}.

\bibitem{Starobinsky:1973aij}
A.~A. Starobinsky,
\newblock Sov. Phys. JETP {\bf 37}, 28 (1973), {\em {Amplification of waves
  reflected from a rotating "black hole".}},
\newblock [Zh. Eksp. Teor. Fiz.64,48(1973)].

\bibitem{1974JETP...38....1S}
A.~A. {Starobinski{\v i}} and S.~M. {Churilov},
\newblock Sov. Phys. JETP {\bf 38}, 1 (1974), {\em {Amplification of
  electromagnetic and gravitational waves scattered by a rotating ``black
  hole''}},
\newblock [Zh. Eksp. Teor. Fiz.65,3(1973)].

\bibitem{Teukolsky:1974yv}
S.~A. Teukolsky and W.~H. Press,
\newblock Astrophys. J. {\bf 193}, 443 (1974), {\em {Perturbations of a
  rotating black hole. III - Interaction of the hole with gravitational and
  electromagnet ic radiation}}.

\bibitem{1972ApJ...178..347B}
J.~M. {Bardeen}, W.~H. {Press}, and S.~A. {Teukolsky},
\newblock \apj {\bf 178}, 347 (1972), {\em {Rotating Black Holes: Locally
  Nonrotating Frames, Energy Extraction, and Scalar Synchrotron Radiation}}.

\bibitem{Penrose:1969pc}
R.~Penrose,
\newblock Riv. Nuovo Cim. {\bf 1}, 252 (1969), {\em {Gravitational collapse:
  The role of general relativity}},
\newblock [Gen. Rel. Grav.34,1141(2002)].

\bibitem{Penrose:1971uk}
R.~Penrose and R.~M. Floyd,
\newblock Nature {\bf 229}, 177 (1971), {\em {Extraction of rotational energy
  from a black hole}}.

\bibitem{Blandford:1977ds}
R.~D. Blandford and R.~L. Znajek,
\newblock Mon. Not. Roy. Astron. Soc. {\bf 179}, 433 (1977), {\em
  {Electromagnetic extractions of energy from Kerr black holes}}.

\bibitem{Guica:2008mu}
M.~Guica, T.~Hartman, W.~Song, and A.~Strominger,
\newblock Phys. Rev. {\bf D80}, 124008 (2009), arXiv:0809.4266[hep-th], {\em {The
  Kerr/CFT Correspondence}}.

\end{thebibliography}
\end{document}